\documentclass[a4paper,11pt]{article}
\usepackage{jheppub} 
\usepackage{lineno}
\usepackage{subcaption} 

\newcommand{\vphi}{\varphi}
\newcommand{\e}{\mbox{e}}


\title{\boldmath Kaluza-Klein monopole with scalar hair}

\author[a]{Y. Brihaye,}
\author[b]{C. Herdeiro,}
\author[b]{J. Novo}
\author[b]{and E. Radu}
\affiliation[a]{Physique-Math\'ematique, Universite de
	Mons-Hainaut, Mons, Belgium}
\affiliation[b]{Departamento de Matem\'atica  da Universidade de Aveiro and
	Center for Research and Development in Mathematics and Applications (CIDMA)
	\\Campus de Santiago, 3810-183 Aveiro, Portugal}

\abstract{
We construct a new family of 
 rotating black holes with scalar hair and a regular horizon
of spherical topology, 
within five dimensional ($d=5$) Einstein’s gravity minimally coupled to a complex, massive scalar field doublet. 
These solutions 
represent generalizations of the Kaluza-Klein monopole  found by Gross, Perry and Sorkin,
with
a twisted $S^1$ bundle over a 
four dimensional Minkowski spacetime being approached in the far field. 
The black holes are described by their mass, angular momentum, tension
and a conserved Noether charge measuring the hairiness of the configurations.
They are supported by rotation and have no static limit,
while for vanishing horizon size, they reduce to boson stars.
When performing a Kaluza-Klein reduction,  
 the $d=5$ solutions yield a family of $d=4$ spherically symmetric
dyonic black holes with gauged scalar hair.
This provides a link between two seemingly unrelated
mechanisms to endow a black hole with scalar hair:
 the $d=5$ \textit{synchronization condition}
between the scalar field frequency and the event horizon angular velocity
results in the $d=4$ \textit{resonance condition}
between the scalar field frequency and the electrostatic 
chemical potential.
}

\numberwithin{equation}{section}

\begin{document}
\maketitle
\flushbottom
\newpage

\section{Introduction}
\label{sec:Intro}

The five dimensional ($d=5$) extension
of Einstein’s theory of General Relativity (GR) was introduced  around one century ago
by
Kaluza \cite{kaluza} 
and Klein \cite{Klein:1926tv}
in an early attempt to unify the
(then) known interactions, namely gravity and electromagnetism.
In the Kaluza-Klein (KK) framework, the Universe has three 
non-compact spatial dimensions;
one extra dimension is compact, 
topologically a circle and sufficiently small as to remain unobservable.

This simple idea has proven to be one of the most fruitful
in theoretical physics,
and
the original KK model has  been extended in various directions, 
such as including more (than one) compact extra-dimensions and starting from higher dimensional theories that are not vacuum GR, including, $e.g.$, gauge and scalar fields -
see~\cite{ACF} for a review and a large set of original literature. 

In the context of this work, a feature of the (initial) KK model is of particular interest:
 the existence of {\it gravitational solitons}, 
$i.e.$  non-singular, horizonless
solutions of the vacuum Einstein field equations.
While no such solutions exist in four spacetime dimensions 
\cite{Einstein:1943ixi,Lichnerowicz,Lichnerowicz1}, 
  gravitational solitons exist in KK theory, as  
found in \textbf{the} '80s by Gross and Perry \cite{Gross:1983hb} and Sorkin \cite{Sorkin:1983ns}.
In the simplest case, this corresponds  to a regular solution, which, 
 from a four-dimensional perspective, describes a (gravitating) Abelian magnetic monopole, 
a feature which has attracted much interest.

The  Gross-Perry-Sorkin (GPS) soliton is the product between the $d=4$ NUT-instanton 
\cite{Hawking:1976jb} 
and a time coordinate, being asymptotically {\it locally} flat only. 
This defines a special type of {\it squashed} KK asymptotics, with 
a twisted $S^1$ bundle over a four dimensional Minkowski spacetime 
being approached in the far field.
 As expected, the vacuum GPS soliton possesses 
 Black Hole  (BHs)  generalizations\footnote{
	There are also BH generalizations of the GPS solution with gauge fields~\cite{Ishihara:2005dp,Yazadjiev:2006iv,Brihaye:2006ws,Nakagawa:2008rm,Nedkova:2012yn}.
} 
 \cite{Chodos:1980df,Dobiasch:1981vh,Pollard:1982gj,Gibbons:1985ac,Wang:2006nw},
with an  event horizon of $S^3$ topology, geometrically being a squashed (rather thand round) sphere.
Such solutions are essentially
higher dimensional near the event horizon,
 but look like four-dimensional - with a compactified
extra-dimension -, at large distances.

\medskip

An interesting question which arises in this context concerns the validity
of the 'no-hair' conjecture \cite{Bekenstein:1996pn}.
In particular, {\it do the BHs with squashed KK asymptotics allow for scalar hair?}
In the last decade it became clear that, at least for asymptotically flat 
\cite{Herdeiro:2014goa}
or  anti-de Sitter \cite{Dias:2011at} 
BHs,
there is a generic mechanism allowing for complex
scalar hair around \textit{rotating}  horizons. 
This mechanism relies on a \textit{synchronization condition} \cite{Herdeiro:2014goa,Herdeiro:2014ima}
which guarantees that there is no scalar energy flux crossing the horizon. 
Mathematically, this results in the following relation between 
the scalar field frequency $\omega$
and 
the event horizon angular velocity
$\Omega_H$:
\begin{eqnarray}
	\label{cond1}
	\omega=m\Omega_H~,
\end{eqnarray}
where $m$ is the winding number which enters the scalar ansatz and
$m \in \mathbb{Z}^+$
for the $d=4$ BHs in \cite{Herdeiro:2014goa,Herdeiro:2014ima}. Eq.~\eqref{cond1} means that 
 the scalar field phase angular velocity matches the horizon's angular velocity;
hence the name 'synchronization'.
This mechanism  appears 
to possess a certain degree of universality, 
applying to both  asymptotically flat  ($d=4$) 
 neutral \cite{Herdeiro:2014goa} and electrically charged \cite{Delgado:2016jxq} rotating BHs, 
as well as to BHs in $d>4$ dimensions
\cite{Brihaye:2014nba,Herdeiro:2015kha},
toroidal horizon topology \cite{Herdeiro:2017oyt},
or 
AdS asymptotics
\cite{Dias:2011at}, 
and even to other spin fields~\cite{Herdeiro:2016tmi,Santos:2020pmh}.

\medskip

One of the main results of this work is to provide evidence that the same mechanism
holds as well for BHs with the same squashed KK asymptotics as the  GPS soliton. 
To do so, we consider $d =5$ Einstein’s gravity minimally 
coupled to a massive complex scalar field doublet,
with a special ansatz, originally introduced in \cite{Hartmann:2010pm},
which factorizes the angular dependence and reduces the problem to solving 
a set of ordinary differential equations (ODEs).
By numerically solving the Einstein-Klein-Gordon
(EKG) equations, we find a four parameter family of regular 
(on and outside the horizon) BHs with scalar hair and squashed KK asymptotics.
The four continuous parameters are 
the mass $M$, 
the angular momentum $J$, 
the tension ${\cal T}$  
and the Noether charge $Q$,  
which measures the scalar field outside the horizon.
For vanishing horizon size, the solutions reduce to solitonic Boson Stars (BSs).
Interestingly, some basic properties of these configurations
are akin to those of the 
$d=4$ BSs and BHs with scalar hair 
\cite{Liebling:2012fv,Herdeiro:2015gia},
rather  than those of the known $d=5$ asymptotically Minkowski EKG solutions 
 \cite{Hartmann:2010pm,Brihaye:2014nba}.

We  also consider the 
equivalent $d=4$ picture, obtained after performing 
a standard KK reduction for both the metric and the scalar field, 
as discussed $e.g.$ in~\cite{Gross:1983hb}.
While the BSs result  in $d=4$ singular configurations - similarly to the dimensional reduction of the (vacuum) GPS monopole -, 
the $d=5$ BHs  
correspond to asymptotically flat  solutions of
a specific $d=4$ Einstein-dilaton-Maxwell-(gauged) scalar (EdMgs) field model.
They are spherically symmetric and describe gravitating dyonic BHs with scalar hair.
Remarkably,
the synchronization condition  (\ref{cond1}) in $d=5$  translates into
the $d=4$  \textit{resonance condition}:
\begin{eqnarray}
	\label{cond2}
	\omega=q_s {\cal V} , 
\end{eqnarray}
which has been found in the study of 
charged (non-spinning) BHs with gauged scalar hair 
\cite{Hong:2019mcj,Herdeiro:2020xmb,Hong:2020miv,Brihaye:2022afz}.
In (\ref{cond2}) 
$q_s$
is the gauge coupling constant, while 
${\cal V} $ 
is the electrostatic chemical potential,  
which, in the $d=5$ picture, corresponds 
to the event horizon angular velocity.

\medskip

This paper is organized as follows.
In Section \ref{section2} we present the EKG model together with a general framework, the ansatz taken - complemented by the equations 
in Appendix \ref{app:Eqs} -,
and discuss the computation of global charges,
together with the solutions of
the Klein-Gordon equation in the  probe limit.
The EKG solutions with
squashed KK asymptotics are discussed in Section \ref{sec:TheSolutions},
where we consider both the case of BSs and BHs.
Section \ref{sec:Asympt} 
is motivated by the observation that 
the GPS soliton
can be taken as an intermediate state between the
five dimensional Minkowski spacetime
and the 'standard' KK vacuum,
$i.e.$ the direct product of
four dimensional Minkowski spacetime  and a circle.
Therefore, in Section~\ref{sec:Asympt}   we consider a comparison between the 
EKG 
solutions in Section \ref{sec:TheSolutions} and those found for the other two spacetime asymptotics 
mentioned above.
In particular, the basic properties of the KK vortices in EKG model are also discussed there
for the first time in the literature.
Section \ref{sec:4Dpic} reconsiders the results from a $d=4$ perspective 
and shows how the $synchronized$ $d=5$ spinning hairy BHs (HBHs) become 
$d=4$ $resonant$ spherically symmetric BHs with gauged scalar hair.
We conclude in Section \ref{sec:Conclusions} with
a discussion and some further remarks. 
A brief review of the vacuum spinning BH solution 
with squashed KK asymptotics 
\cite{Dobiasch:1981vh,Wang:2006nw} 
is presented in Appendix \ref{app:AppVacBH}, 
as well as an exact solution of the KG equation 
on an extremal (vacuum) BH background.

\section{The framework}
\label{section2}

\subsection{Action and field equations}
\label{section21}

We consider the $d=5$ Einstein's gravity minimally coupled
to a massive complex scalar field doublet $\Psi$, 
with action  
\begin{equation}
	\label{action}
	\mathcal{S}= \frac{1}{4 \pi G_5}\int_\mathcal{M}  d^5x \sqrt{-g}\left[ \frac{1}{4}R
	-\frac{1}{2} g^{ab}\left( \Psi_{, \, a}^\dagger \Psi_{, \, b} + \Psi _
	{, \, b}^\dagger \Psi _{, \, a} \right) 
-\mu^2 \Psi^\dagger \Psi  
	\right] 
	-\frac{1}{8 \pi G_5}\int_{\partial \mathcal{M}} d^4 x\sqrt{-h}K,
\end{equation}
where
 $\dagger$ denotes the complex transpose,
 $G_5$ is the five dimensional Newton's constant, which will be set to unity in the  numerics,
 $\mu$ is the scalar field mass,  
  $h_{ij}$ is the induced metric on the boundary $\partial \mathcal{M}$ of the spacetime $\mathcal{M}$,
and $K_{ij}$ is the extrinsic curvature of this boundary, with $K=K_{ij}h^{ij}$.

Variation of this action with respect to the metric and scalar
field gives the EKG equations:
\begin{equation}
	\label{EKG-eqs}
	R_{ab}-\frac{1}{2}g_{ab}R = 2 ~T_{ab}\,,
	~~\left(
	\Box
	 -\mu^2 \right)\Psi=0~,
\end{equation}  
where 
\begin{equation}
	\label{Tab}
	T_{ab}= 
	\Psi_{ , a}^\dagger\Psi_{,b}
	+\Psi_{,b}^\dagger\Psi_{,a} 
	-g_{ab}  \left[ \frac{1}{2} g^{cd} 
	\left( \Psi_{,c}^\dagger\Psi_{,d}+
	\Psi_{,d}^\dagger\Psi_{,c} \right)
	+\mu^2 \Psi^\dagger \Psi
	\right]~,
\end{equation}
is the
stress-energy tensor  of the scalar field.

%
%

\subsection{The vacuum  Gross-Perry-Sorkin solution  
}
\label{sec:Background}

We start by introducing the squashed Kaluza-Klein (KK) geometry
found in~\cite{Gross:1983hb,Sorkin:1983ns},
which captures some of the basic features
 of the solutions constructed in this work.
This metric
solves the vacuum Einstein equations, $\Psi=0$,
and
is the product between the $d=4$
(self-dual) Euclidean Taub-NUT instanton 
\cite{Hawking:1976jb} 
and a time coordinate,
\begin{equation}
	\label{metric-gen}
	ds^2=-dt^2+ds_{4}^2\ .
\end{equation}  
The instanton metric 
$ds_{4}^2$
possesses an intrinsic length scale $N \geqslant 0$,
which is an input parameter: the NUT charge.
The geometry can be written with
several different choices of the radial coordinate $r$,
which make more transparent  
various limits of interest. 

The first form for the instanton metric we shall consider is
\begin{eqnarray}
	\label{metric1}
	ds_{4}^2=\frac{r+N}{r-N} dr^2
	+ (r^2-N^2) (d\theta^2+\sin^2 \theta d\varphi^2)
	+\frac{r-N}{r+N} 4N^2 (d\psi+\cos \theta d\varphi)^2 .
\end{eqnarray}
The range of the radial coordinate
is $N\leqslant r< \infty$,
while 
$\theta ,\, \varphi$ and $ \psi  $  
are the usual Euler angles with ranges
$0\leqslant \theta \leqslant \pi$,
$0\leqslant \varphi<2\pi $,
$0 \leqslant \psi<4\pi$. 
This metric is asymptotically \textit{locally} flat,
in the sense that 
the curvature goes to zero as $r\to \infty$.
The surfaces of constant $r$
are topologically $S^3$, 
although their metric is a deformed 3-sphere, with an $S^1$ fiber over $S^2$.
Also, $r=N$ corresponds to the origin of the coordinates,
on $\mathbb{R}^4$, 
with
the size of $S^3$ shrinking to zero.

The coordinate transformation
\begin{eqnarray}
	r\to r-N
\end{eqnarray}
leads to an equivalent form of (\ref{metric1}), 
\begin{eqnarray}
	\label{metric2}
	ds_{4}^2=\left(1+\frac{2N}{r}\right)
	\left[ dr^2
	+  r^2  (d\theta^2+\sin^2 \theta d\varphi^2)
	\right]
	+\frac{4N^2}{1+\frac{2N}{r}}  (d\psi+\cos \theta d\varphi)^2 , 
\end{eqnarray}
but now with the usual range of the new radial coordinate,
$0\leqslant r< \infty$.

The $N\to 0$ limit  of the $d=4$ NUT instanton 
corresponds to the flat $\mathbb{R}^3 \times S^1$
space. To take this limit,  
one defines a new coordinate
\begin{eqnarray}
	\label{z}
	\psi=\frac{z}{2N}\,.
\end{eqnarray}
Then, as 
$N\to 0$, 
one finds the line element 
\begin{eqnarray}
	\label{metricM4S1}
	ds^2=-dt^2+dr^2+r^2(d\theta^2+\sin^2 \theta d\varphi^2)+dz^2 \ ,
\end{eqnarray}
which is the  $\mathbb{M}^{1,3} \times S^1$ metric, 
with an arbitrary periodicity $L$
for the coordinate $z$.

An alternative form of the instanton metric,
which shall later be employed in the construction of BSs, 
is obtained by taking the following coordinate transformation
in (\ref{metric1})
\begin{eqnarray}
	\label{transf1}
	r \to N+\frac{r^2}{8N}\,,
\end{eqnarray}
with the new radial coordinate ranging from zero to infinity.
This result in the line element
\begin{equation}
	\label{metric3}
	ds_{4}^2= 
	\left(1+\frac{r^2}{16 N^2} \right)
	\left[ dr^2
	+  \frac{r^2 }{4} 
	(
	d\theta^2+\sin^2 \theta d\varphi^2 
	)
	\right]
	+\frac{1}{4}\left(
	\frac{r^2}{1+\frac{r^2}{16 N^2}} \right)
	(d\psi+\cos \theta d\varphi)^2 
	. 
\end{equation} 
Then the limit 
$N \to \infty$  corresponds to 
\begin{eqnarray}
	\label{metricM5}
	ds =-dt^2+ 
	dr^2+\frac{r^2}{4}
	\left[ 
	d\theta^2+\sin^2 d\varphi^2+(d\psi+\cos \theta d\varphi)^2
	\right] \ ,
\end{eqnarray}
which is the $d=5$ Minkowski spacetime $\mathbb{M}^{1,4}$.
 
As such, 
when varying the  parameter $N$,
one can consider the $d=5$ metric (\ref{metric-gen})
as interpolating
between
the 'standard' KK vacuum, $i.e.$ the direct product of
$d=4$ Minkowski spacetime  and a circle -- 
the limit $N \to 0$ --,
and the
$d=5$ Minkowski spacetime   --
the limit $N\to \infty$.

As expected, the GPS soliton
possesses BH generalizations, with
a squashed horizon of $S^3$ topology
and nonzero size, 
whose basic properties are reviewed in 
Appendix A.

\subsection{A general ansatz 
}


The geometries studied in this work 
are described by a generic line element\footnote{There is a residual metric gauge freedom in (\ref{metric}), to be fixed later.
The GPS metric (\ref{metric-gen}), with the various choices of radial coordinate in the spatial part,
is of the form  (\ref{metric}).}
\begin{eqnarray}
	\label{metric}
	ds^2&=& - {\cal F}_0(r)dt^2+{\cal F}_1(r)  dr^2
	+ {\cal F}_2(r)
	\left(
	\sigma_1^2+\sigma_2^2
	\right)
	+ {\cal F}_3(r)
	(\sigma_3-2W(r)dt)^2
	\ ,
\end{eqnarray}
with the left-invariant 
1-forms $\sigma_i$ on $S^3$, 
\begin{eqnarray}
	\nonumber
	&&\sigma_1=\cos  \psi d  \theta+\sin \psi \sin  \theta d \varphi,
	\\
	\label{sigma}
	&&\sigma_2=-\sin  \psi d  \theta+\cos \psi \sin   \theta d  \varphi,
	\\
	\nonumber
	&&\sigma_3=d \psi  + \cos  \theta d  \varphi,
\end{eqnarray}
and $(\theta , \varphi , \psi)$ the usual Euler angles defined above, 
while  $r$ and $t$ denote the radial and time coordinates, respectively\footnote{
 An equivalent form of this line element (used sometimes in the literature)
is found by defining the new coordinates
$\Theta= \theta/2 $,
$\varphi_1=(\psi-\varphi)/2$,
$\varphi_2=(\psi+\varphi)/2$
(with 
$0\leqslant \Theta \leqslant \pi/2$,
$0\leqslant (\varphi_1,\varphi_2)< 2\pi $),
yielding
\begin{eqnarray}
	\label{metricalt}
		ds^2=- {\cal F}_0(r) dt^2+ {\cal F}_1(r)  dr^2
	+ 4 \bigg[ 
 {\cal F}_2(r)  d\Theta^2
 	+{\cal F}_3(r)  (
 \sin^2\Theta  (d\varphi_1-W dt)^2
 	+	\cos^2\Theta  (d\varphi_2-W dt)^2
 								)
								\bigg] \ \ \ \ 
								\\
								\nonumber
 +[{\cal F}_2(r)-{\cal F}_3(r) ]\sin^2(2\Theta)	(d\varphi_1-	d\varphi_2)^2						
 	\ . \ \ \ 
	\end{eqnarray}
}.
Apart from the Killing vector $K_0=\partial_{\psi}$, the line element (\ref{metric}) possesses
three additional Killing vectors 
\begin{eqnarray} 
\nonumber
	K_1 = \sin \varphi \partial_{\theta}\,,\quad 
	K_2 =  -\cos \varphi \partial_{\theta}
	+\sin \varphi \cot \theta  \partial_{\varphi}
	-\frac{\sin\varphi}{\sin\theta}  \partial_{\psi}\,,\quad 
	K_3 =\partial_{\varphi}\,,
\end{eqnarray} 
which obey an $SU(2)$ algebra.

Concerning the scalar sector, we shall consider a general ansatz, 
with 
\begin{equation}
\label{scalarS}
	\Psi=\phi\left(r\right)e^{-i\omega t}\hat{\Psi}_s\,,\quad {\rm with}~~~s=0,1\,,
\end{equation}
where
\begin{eqnarray}
\label{scalar}
	\hat{\Psi}_0=\left( \begin{array}{c} 
		1
		\\
		0 
	\end{array} \right)\,,\quad \hat{\Psi}_1=\left( \begin{array}{c} 
	~\sin\frac{\theta}{2}~e^{-i \frac{\vphi }{2}} 
	\\
	\cos\frac{\theta}{2}~e^{i \frac{\vphi }{2}}
\end{array}\right)e^{ i \frac{ \psi}{2}}\,.
\end{eqnarray}
The case $s=0$ is only compatible with a static line-element, 
 in which case 
similar results are found when considering a  singlet scalar field, $i.e.$ $\Psi=\phi(r)e^{-i\omega t}$.
In the rotating case, we shall use instead the 
$s=1$
scalar ansatz, which was
originally proposed in~\cite{Hartmann:2010pm}, albeit  
for a parametrization of the 3-sphere 
in terms of $\{ \Theta,\varphi_1,\varphi_2 \}$ - 
see Eq. (\ref{metricalt}).

\medskip
Both solitons and BHs, 
can be studied  by using the general  metric form (\ref{metric})  
together with the scalar ansatz (\ref{scalarS}), (\ref{scalar}).
The  corresponding expressions of the Einstein
and energy-momentum 
tensors are given in Appendix A; the resulting EKG equations depend only on the radial
variable $r$.
The BHs have a regular horizon located at some $r_H >0$,
with ${\cal F}_0(r_H)=0$.
For solitons, the horizon is replaced with a regular origin $r=0$, 
where both 
${\cal F}_2$
and 
${\cal F}_3$
vanish, while 
${\cal F}_0$ and 
${\cal F}_1$ are finite and nonzero.
In both cases, the solutions share the far field asymptotics with
the GPS metric 
(\ref{metric-gen}), 	
(\ref{metric1}),
with
${\cal F}_0 \to 1$,
${\cal F}_1 \to 1$,
${\cal F}_2 \to r^2$,
${\cal F}_3 \to 4N^2$,
$W\to 0$ 
(and also  $\phi\to 0$)
as $r\to \infty$.
 
Finally, let us mention that
the action (\ref{action}) is invariant under the {\it global} \ $U(1)$ transformation
$\Psi\rightarrow e^{i\alpha}\Psi$, where $\alpha$ is a constant.
This implies that the current 
$j^a=-i (\Psi^* \partial^a \Psi-\Psi \partial^a \Psi^*)$
is conserved, $i.e.$ $j^a_{\ ;a}=0$.
Therefore integrating the timelike component of this current on a spacelike slice $\Sigma$ yields a conserved quantity -- the \textit{Noether charge}:
\begin{eqnarray}
	\label{Q}
	Q=\int_{\Sigma}~j^t = 32 \pi^2 \int_{r_0}^\infty dr~ F_2 \sqrt{\frac{F_1 F_3}{F_0}}
	(\omega-W) \phi^2,
\end{eqnarray}
where $r_0= \{0,r_H \}$ for solitons and BHs respectively.

\subsection{The computation of global changes}
\label{charges}

Apart from the Noether charge, the solutions possess three more
conserved quantities: mass $M$, angular momentum $J$ 
and tension\footnote{  The tension of a spacetime was first introduced 
in \cite{Traschen:2001pb,Townsend:2001rg}. 
This global charge is associated with the translation symmetry along the  extra-dimension, 
in a similar way to the mass being related with the existence of a timelike Killing vector field.}
 ${\cal T}$,
whose values are encoded in the far field form of the metric functions.
Given the non-standard asymptotics of the solutions in this work, 
one way to compute their charges 
is to use the
quasilocal tensor of Brown and York \cite{Brown:1992br}, augmented by the counterterm 
formalism 
\cite{Kraus:1999di,Lau:1999dp,Mann:1999pc,Astefanesei:2005ad}.
This technique, inspired by the holographic
renormalization method in spacetimes with anti-de Sitter (AdS) asymptotics 
\cite{Henningson:1998gx,Balasubramanian:1999re}
 consists in adding a suitable boundary counterterm  $\mathcal{S}_{\rm ct}$ to the action of the theory;
 thus the  
bulk equations of motion are not altered. 
$\mathcal{S}_{\rm ct}$ is built up with curvature invariants of the induced metric on the boundary, which is sent to spatial infinity after the integration. 
Unlike the background substraction method (see below), this procedure  
is intrinsic to the spacetime of interest and it is unambiguous once 
the counterterm is specified.
In our case, however, differently from the AdS case, this method has the drawback that
there is  no rigorous justification for the choice of  the counterterm.

In this work  
we shall use the counterterm proposed in~\cite{Mann:2005cx} to compute the mass of the KK monopole,
with 
\begin{equation}
	\label{countertermaction}
	\mathcal{S}_{\rm ct}=\frac{1}{8\pi G_5}\int_{\partial \mathcal{M}} d^{4}x\sqrt{-h}\sqrt{2\mathsf{R}} \,
	, 
\end{equation}
where $\mathsf{R}$ is the Ricci scalar of the induced metric on
the boundary.  
The variation of this action $w.r.t.$ $h_{ij}$
results in the  boundary
stress-energy tensor 
\begin{equation}
	\label{mann} T_{ij}=\frac{1}{8\pi G_5}\left( K_{ij}-Kh_{ij}-\Phi(
	\mathsf{R}_{ij}-\mathsf{R}h_{ij})-h_{ij}h^{kl}\Phi_{;kl}+\Phi_{;ij}
	\right)\,,
\end{equation}
where  we defined $\Phi=\sqrt{2/{\mathsf{R}}}$.
If the boundary
geometry has an isometry generated by a Killing vector $\xi ^{i}$,
then $ T_{ij}\xi ^{j}$ is divergence free, 
from which it follows
that the quantity
\begin{equation}
	\mathcal{Q}=\int_{\Sigma }d\Sigma_{i}T^{i}{}_{j}\xi ^{j},
	\label{concharge}
\end{equation}
associated with a closed surface $\Sigma$, is conserved. 
Physically, this means that the observers
on the boundary with the induced metric $h_{ij}$ 
measure the same value of $\mathcal{Q}$.
For the considered framework, the mass $M$, tension ${\cal T}$
and angular momentum\footnote{When considering the coordinates of eq. (\ref{metricalt}), 
this angular momentum is related to equal
rotations $w.r.t.$ the angular directions $\varphi_1$ and $\varphi_2$.}
 $J$
are computed as the integrals of the
boundary
stress-energy tensor 
components
$T_t^t$,
$T_\psi^\psi$
and
$T_\psi^t$, 
respectively.
Interestingly, as found in  \cite{Mann:2005cx},
this approach predicts a non-zero value for the
mass and tension of the GPS soliton (\ref{metric-gen}), respectively
\begin{eqnarray}
\label{MTvac}
	M=M_0=\frac{4\pi N^2}{G_5},~~{\cal T}={\cal T}_0=-\frac{N}{G_5}~,
\end{eqnarray}
while the angular momentum is zero, as expected. 
When considering  solutions of the EKG equations with the same asymptotics,
$M$ and ${\cal T}$
 will contain the above background contributions.


Apart from the boundary counterterm method,
we have also computed $M,{\cal T}$ and $J$ by using the Abbott-Deser approach \cite{Abbott:1981ff}. 
This was initially proposed to address 
the issue of conserved charges of asymptotically (anti-)de Sitter  spacetime,
 but has also proved useful for other asymptotic behaviors.
 In this approach, conserved charges are associated with isometries of some background geometry
and can be summarized as follows. 
First, the following decomposition of the metric is introduced
\begin{equation}
	g_{ab}=\bar{g}_{ ab}+ \bar h_{ab}\,,
\end{equation}
where $\bar{g}_{ab}$ corresponds to the {\it background metric} and $\bar h_{ab}$ is a perturbation.
Assuming that $\bar{g}_{ab}$ is a vacuum solution, then the field equations can be written as:
\begin{equation}
	R^{ab}_{L}-\frac{1}{2}\bar{g}^{ ab}R_{L}=\tilde{T}^{ab}\,,
\end{equation}
where the subscript $L$ denotes terms that are linear in the perturbation and
  $\tilde{T} ^{ab}$ collects all higher order terms in $\bar h$ 
(the tilde serves to distinguish
 it from the boundary
stress-energy tensor defined above for the counterterm method). 
It can be shown that the left-hand side of the above equation 
satisfies the Bianchi identity $w.r.t.$ the background metric 
- see $e.g.$ chap. 6 of \cite{Ortin:2004ms};
 then the field equations imply:
\begin{equation}
	\bar{\nabla}_a \tilde{T}^{a b}=0\, ,
\end{equation}
where the bar indicates that the covariant derivative 
is taken $w.r.t.$ the background metric. 
If $\chi$ is a Killing vector of the background geometry,  
the following conservation law holds:
\begin{equation}
	\bar{\nabla}_a\left(\tilde{T}^{a b}\chi_b\right)=0\, ,
\end{equation}
which allows  to define the associated conserved charge:
\begin{equation}
	\tilde{\mathcal{Q}}=\int_{\Sigma} d\Sigma_{i} \tilde{T}^{ij} \chi_j \,,
\end{equation}
where $\Sigma$ is a closed surface.

The Abbott-Deser method has been extensively 
used to compute conserved charges on some non-asymptotically flat spacetimes,
see $e.g.$~\cite{Cai:2006td,Wang:2006nw,Peng:2017bjo}
for the case of
  KK asymptotics.
One should, however, mention that 
the choice  of the background metric $\bar{g}$
requires special care. 
It is sometimes common to consider the asymptotic metric as the background;
however this is not always a solution to Einstein's equations, 
in which case the effective energy-momentum tensor $\tilde{T}^{\mu\nu}$ 
has contributions not only from the perturbation, but also from the background. 
For asymptotically squashed KK spacetimes,
this issue 
has been explored in~\cite{Peng:2017bjo}.
The results there show that,  
 when choosing the asymptotic form of the
metric (\ref{metric-gen}), (\ref{metric1}), 
 as the background\footnote{ 
The large$-r$ limit of the GPS metric
  (\ref{metric-gen}), (\ref{metric1}), 
	results in the  line element
\begin{eqnarray}
\label{GPSinf}
 ds^2=-dt^2+dr^2+r^2(d\theta^2+\sin^2 \theta d\varphi^2) +4N^2(d\psi+\cos \theta d\varphi)^2, 
\end{eqnarray}
 which, however,  {\it does not} solve the vacuum Einstein equations.},
 the mass and tension of the GPS soliton take the same (nonzero) values (\ref{MTvac})  
as found for the counterterm approach.  

\medskip
The global charges of the solutions in this  work 
were computed using both methods described above.
We have found that
 the values of $M$, ${\cal T}$ and $J$ obtained within
 the counterterm approach and Abbott-Deser approach 
- with the asymptotic metric (\ref{GPSinf}) 
taken as the reference background -, agree\footnote{
As expected, however, when using the Abbott-Deser approach
with  the GPS metric as background, the terms  $M_0$ and ${\cal T}_0$, $cf.$ eq. (\ref{MTvac}),
are absent in the expression of 
  mass and tension of the EKG solutions.}. 
But in order to simplify the picture, and in particular
to make clear the limit of
a vanishing scalar field,
we have subtracted the $M_0$-term in all figures where 
the mass of solutions with squashed KK asymptotics is displayed, which is nonetheless in the 
 corresponding equations.

\subsection{The probe limit: no scalar clouds   }
\label{section13}

Before discussing the  solutions of the full
system (\ref{EKG-eqs}),
it is  of interest to consider the solutions of the KG equation 
in the probe-limit case,
$i.e.$ ignoring the backreaction on the spacetime geometry.

Starting with the case of a GPS background 
as given by
 (\ref{metric-gen})  and (\ref{metric1}), 
the KG equation   reads 
(where a prime denotes the derivative $w.r.t.$ the radial coordinate $r$):
\begin{equation}
	\label{KGeq-gen}
		\phi^{\prime\prime}+  \frac{2\phi'(r)}{r-N}+\frac{ r+N}{r-N}	
			\left(\omega^{2}-\mu_{\rm eff}^{2}\right)\phi - \frac{s (r+5N)}{8N(r-N)^2}\phi=0,
\end{equation}
with $s=0,1$
for the scalar ansatz 
(\ref{scalarS}),
  (\ref{scalar}).
Also, in the above expression we define  
\begin{equation}
\label{mueff}
	\mu_{\rm eff}= \sqrt{\mu^2+\frac{s}{16N^2}}\,.
\end{equation}
That is, for $s=1$, the scalar field acquires a contribution 
to the mass coming from the dependence on the $\psi$-direction, 
giving rise to an \textit{effective mass}, 
a features shared also by the gravitating
solutions. 
The bound state condition then needs to consider this effective mass, 
$i.e.$ $\omega^2\leqslant \mu_{\rm eff}^2$.

For the scalar Ansatz 
(\ref{scalarS}),
 (\ref{scalar}) with $s=0$,
the general solution of the equation (\ref{KGeq-gen}) 
reads
$\phi(r)=c_1 \phi_1(r) + c_2 \phi_2(r)$
where
$c_1,c_2$ are arbitrary constants, and
\begin{eqnarray}
	&&
	\label{solKK1}
        \phi_1(r) =\frac{N}{r-N}\e^{-\left(r-N\right)}U\left(N \sqrt{\mu^2 -\omega^2 },0,2 (r-N) \sqrt{\mu^2 -\omega^2}\right)
	\\
	&&	
	\nonumber
         \phi_2(r) =\frac{N}{r-N}\e^{-\left(r-N\right)}L_{-N \sqrt{\mu^2 -\omega^2}}^{-1}\left(2 (r-N) \sqrt{\mu^2 -\omega^2}\right)
\end{eqnarray}
Here
$U$ is the confluent hypergeometric function and 
$L$ is the generalized Laguerre polynomial. 
For solutions satisfying the
bound state condition, 
the function
$\phi_1(r)$ diverges at $r=N$ while $\phi_2(r)$ diverges at infinity\footnote{
	In the limiting case with $\omega^2=\mu^2$, the   solution (\ref{solKK1})
	takes the simpler form
	$\phi(r)=\frac{c_1}{r-N}+c_2\,$.
}.

For the  scalar ansatz 
(\ref{scalarS}),
(\ref{scalar}) 
with an explicit dependence of angular coordinates
($i.e.$ $s=1$),
the general solution is
$\phi(r)=c_1 \phi_1(r) + c_2 \phi_2(r)$ , where
\begin{align}
\label{spinKK}
	\phi_1(r) &= \sqrt{\frac{r}{N}-1} 
	e^{-(r-N) \sqrt{\mu_{\rm eff}^2-\omega^2}} U\left( \frac{c}{2} 3,	2 (r-N)\sqrt{\mu_{\rm eff}^2-\omega^2}\right)
        \,,
	\\
	\nonumber
	\phi_2(r) &=  \sqrt{\frac{r}{N}-1} 
	e^{-(r-N) \sqrt{\mu_{\rm eff}^2-\omega^2}} 
        L_{-\frac{c}{2}}^2\left(2 (r-H) \sqrt{\mu_{\rm eff}^2-\omega
   ^2}\right)
\end{align}
with 
$c=3+1/{(8N\sqrt{\mu_{\rm eff}^2-\omega^2})}+2N \sqrt{\mu_{\rm eff}^2-\omega^2}$,
which is again divergent at $r=N$ or at infinity\footnote{
	In the special case with $\omega^2=\mu_{\rm eff}^2$, the general solution 
	(\ref{spinKK})
	simplifies to
	\begin{equation}
		\phi(r)= \frac{4\sqrt{2N}}{\sqrt{r-N}}
		\left[
		c_1 I_2\left( \sqrt{\frac{r-N}{2N} }\right)+	c_2 K_2\left(\sqrt{\frac{r-N}{1N} }\right)
		\right],
		\label{eq:margSol}
	\end{equation}
	where $I$ and $K$ are Bessel functions, 
	which is again divergent.}.
Therefore we conclude that there are no scalar clouds on a GPS soliton background, 
a situation  also found for the 
 $\mathbb{M}^{1,3} \times S^1$  
or $\mathbb{M}^{1,4}  $ cases. 

In $d=4$,  (real frequency) bound states 
are found for a particular set of Kerr BHs.
These configurations are at the threshold of the superradiant instability, 
the scalar field satisfying  the synchronization condition 
(\ref{cond1})~\cite{Hod:2012px,Hod:2013zza}.  
 Inspired by this result, 
we have looked for scalar clouds on a
spinning vacuum BH with squashed KK asymptotics, 
whose metric is presented in Appendix \ref{sec:AppRotBH},
the scalar field ansatz  being the case $s=1$ in (\ref{scalar}).
Unfortunately,  in the generic, non-extremal case,
 the resulting equation for the scalar amplitude $\phi(r)$ could not be solved analytically.
Therefore we have considered a numerical approach, 
employing similar methods as the ones described   $e.g.$ in~\cite{Benone:2014ssa}.
 After imposing the synchronization condition, Eq. (\ref{cond1}),
 we looked for scalar configurations which are regular on and outside the horizon
and vanish at infinity. 
We found no indication that such solution exist, despite not being able to provide a non-existence proof. 
This result is also supported by the exact solutions
 (\ref{sol-extremal})
 found for the extremal (spinning) BH background,
which is displayed in Appendix \ref{sec:AppRotBH}. 
Unlike the case of an extremal Kerr background in $d=4$~\cite{Hod:2012px},
	this solution is singular at the horizon or at infinity. 
 
This non-existence result implies the absence
of an {\it existence line}
for the HBHs with squashed KK asymptotics,
which would be given by the set of vacuum BH configurations
allowing for {\it scalar clouds}. 
We mention that a similar result
\cite{Cardoso:2005vk,Kunduri:2006qa}
 is found for
for a test massive scalar field  on the background of an 
asymptotically $\mathbb{M}^{1,4}$
Myers-Perry BHs 
\cite{Myers:1986un}.

\section{The solutions
}
\label{sec:TheSolutions}

\subsection{The numerical approach
}
The numerical  methods employed here are similar to those used in~\cite{Brihaye:2014nba}
to study $d=5$ EKG solutions with equal-magnitude angular momenta and
$\mathbb{M}^{1,4}$ asymptotics.
The BH problem possesses four input parameters  (we recall that $G_5=1$):
two of them belong to a specific model -- the scalar field mass $\mu$ and the NUT parameter $N$; 
and two specify a solution -- the field frequency $\omega$ and the horizon radius $r_H$
(with $r_H=0$ for solitons).
In practice,
dimensionless variables and global quantities are introduced by using
natural units set by the scalar field mass $\mu$, 
$e.g.$ 
$r\to r/\mu$,
$\omega \to \omega/\mu$
and
$N \to N \mu$, 
which reduce to taking $\mu=1$ in the input of the numerical code\footnote{In principle,
the effective mass relation for the scalar field, Eq. (\ref{mueff}),
would allow for the existence of spinning solutions with $\mu=0$.
However, so far these could not be obtained. }. 
As such, we are left with three (two) input parameters for BHs (solitons).

The system of five non-linear coupled ODEs
for the metric functions
 and the scalar amplitude, subject to the  boundary conditions described below,
was solved using two different solvers.
The BH solutions and a part of the BS sets were found
by using a professional package  
based on the  Newton-Raphson method
\cite{schoen}.
In this case we have introduced a compactified coordinate 
$x$, where $0\leqslant x\leqslant 1$;
the relation between
the usual radial coordinate $r$
and $x$  
being $r=(r_H+c x)/(1-x)$, with $c$ a suitable chosen constant, usually of order one.
Typical grids used have around $800$ points,
distributed equidistantly in $x$. 

Most of the BSs solutions were found
by using another package,
which employs a collocation
method for boundary-value ordinary differential equations and a damped Newton method
of quasi-linearization  \cite{COLSYS}. 
The meshes here are non-equidistant and use around 300 points in the interval $0\leqslant r<r_{\rm max}$,
with $r_{\rm max}$ around $10^{4}$. 

We have compared a number of BS solutions 
constructed with these two different methods
and found a very good agreement between them. 
In both cases, the
constraint Einstein equation
and
the Smarr relation (\ref{Smarr}) 
have been used to test 
the accuracy of the results.
Based on that, we estimate a typical relative error $<10^{-5}$
for the solutions reported herein.
The numerical accuracy, however, decreases close to the 
maximal value of the frequency and also for solutions close to the center
of the $(\omega,M)$-spiral - see below.

Finally, let us mention that 
in this work  we report results for a
nodeless scalar field amplitude only,
although solutions with nodes exist as well.

\subsection{Boson Stars}

Before discussing the BH solutions, 
it is useful to first describe the properties of their solitonic limit, 
$i.e.$ of the BSs.  
In the numerical construction of these horizonless solutions,
we have found useful to use the following metric ansatz\footnote{
Since 
$g^{tt}=-e^{-2F_0(r)}<0$, the $t-$coordinate provides a
global time function and the spacetime is free of causal pathologies, which is also the case for BHs.}
%
\begin{figure}[htbp]
	\makebox[\linewidth][c]{%
		\begin{subfigure}[b]{8cm}
			\centering
			\includegraphics[width=8cm]{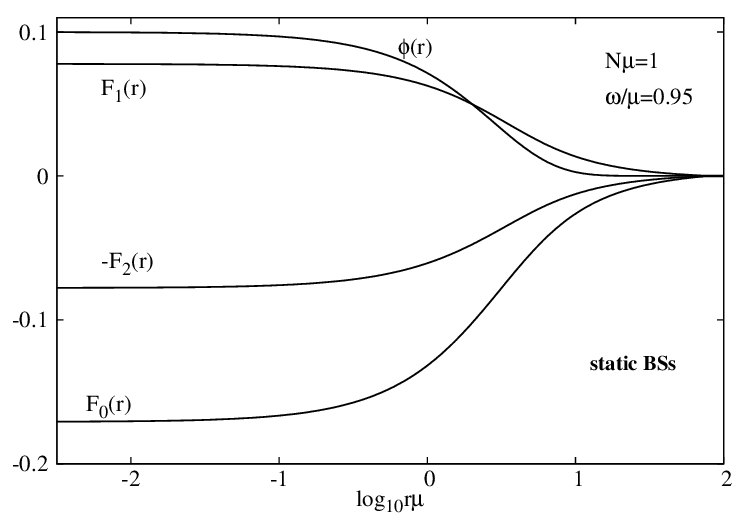}
		\end{subfigure}%
		\begin{subfigure}[b]{8cm}
			\centering
			\includegraphics[width=8cm]{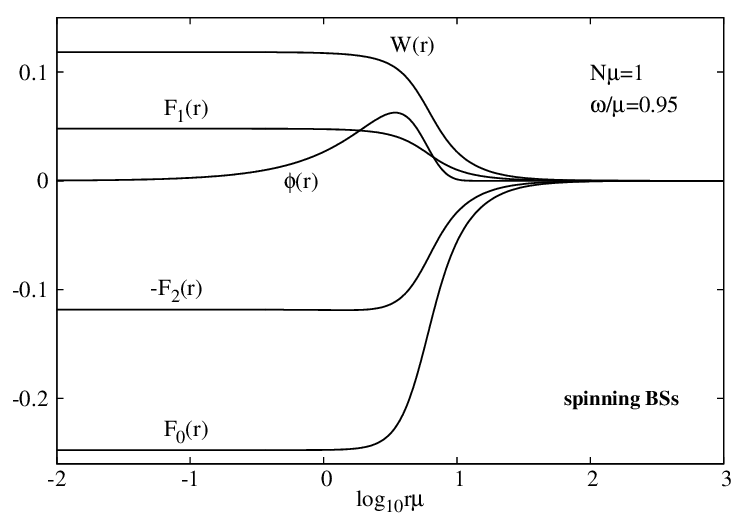}
		\end{subfigure}%
	}
	\caption{
	{\small
	The profile of a typical static (left panel)
		and spinning (right panel) BS 
		with the same  input parameters.
		}
		\label{fig:1}
		}
\end{figure}
- which can be viewed as a deformation
of  (\ref{metric3}) -, in terms of four unknown functions $(F_0,F_1,F_2,W)$:
\begin{align}
	\nonumber
	ds^2=  -e^{2F_0(r)} dt^2&+e^{2F_1(r)} \left(1+\frac{r^2}{16 N^2} \right)
	\bigg[
	dr^2+\frac{1}{4}r^2\left(\sigma_1^2+ \sigma_2^2 \right)
	\bigg]
	\\
	\label{metricBS}
	&+\frac{e^{2F_2(r)}}{4}\left(\frac{r^2}{1+\frac{r^2}{16 N^2}}\right)
	\left(\sigma_3-2W(r) dt\right)^2 \, .
\end{align} 
 
Near the origin ($r=0$), the solutions possess a formal
power series expansion, whose first terms are given by
\begin{eqnarray}
	&&
	\label{r0BS}
	F_0(r)=f_{00}^{(0)}+f_{02}^{(0)}r^2+\dots,~~
	F_1(r)=f_{10}^{(0)}+f_{12}^{(0)}r^2+\dots,~~
	\\
	&&
	\nonumber
	F_2(r)=f_{10}^{(0)}+f_{22}^{(0)}r^2+\dots,~~
	W(r)=w_0+w_{2}^{(0)}r^2+\dots,~~
	\phi(r)=\phi_1 r+\phi_3 r^3+\dots,~~
\end{eqnarray}
with all coefficients  fixed by 
$f_{00}^{(0)}$, 
$f_{10}^{(0)}$, 
$f_{22}^{(0)}$,
$w_0$
and 
$\phi_1$,
$e.g.$ 
$f_{12}^{(0)}=-f_{22}^{(0)}-2\phi_1^2/3$. 

For the far field expansion of the solutions, one finds
\begin{eqnarray}
	&&
	\nonumber
	F_0(r)=\frac{f_{02}}{r^2}+\dots,~~
	F_1(r)=\frac{f_{12}}{r^2}+\dots,~~
	F_2(r)=-\frac{(f_{02}+f_{12})}{r^2}+\dots,~~
	\\
	&&
	\label{infBS}
	W(r)=\frac{w_{2}}{r^2}+\dots,~~
	\phi(r)=c_1\frac{e^{-r^2\sqrt{\mu^2_{\rm eff}-\omega^2}}}{r^2}+\dots,~~
\end{eqnarray}
with the parameters 
$f_{02}$,
$f_{12}$,
$w_2$
and $c_1$
being fixed by numerics.

The mass, tension and angular momentum are defined in terms of the asymptotic coefficients by
(note the presence of the background terms (\ref{MTvac}) in $M,{\cal T}$) 
\begin{eqnarray}
	M= \frac{4 \pi }{G_5}\left[N^2+\frac{1}{8}(f_{12}-f_{02}) \right], ~
	{\cal T}=-\frac{1}{G}_5\left[N +\frac{1}{4N}(f_{12}+\frac{1}{2}f_{02}) \right], ~
~
	J=\frac{2 \pi N^2 w_2 }{G_5}\,.~~{~~~}
\end{eqnarray} 
One can easily show that, as with 
BSs with other asymptotics,
 the Noether charge and the angular momenta of the spinning BSs 
are not independent quantities,
with
\begin{eqnarray}
	\label{QJ}
	Q=2J \,.
\end{eqnarray}

\begin{figure}[htbp]
	\makebox[\linewidth][c]{%
		\begin{subfigure}[b]{8cm}
			\centering
			\includegraphics[width=8cm]{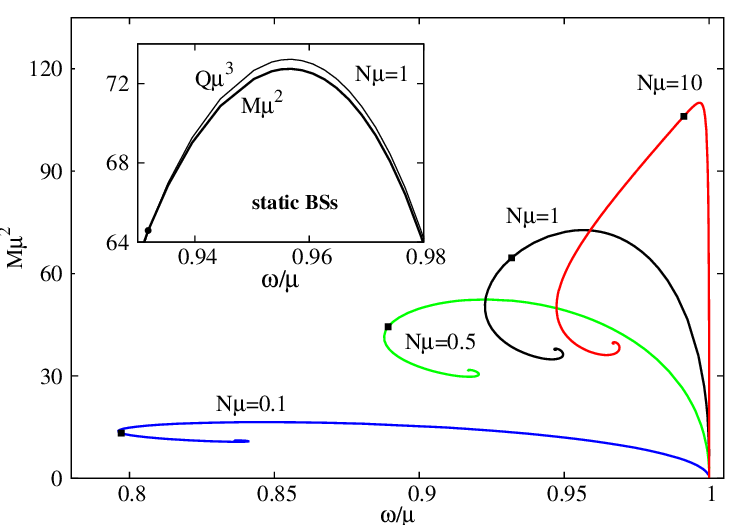}
		\end{subfigure}%
		\begin{subfigure}[b]{8cm}
			\centering
			\includegraphics[width=8cm]{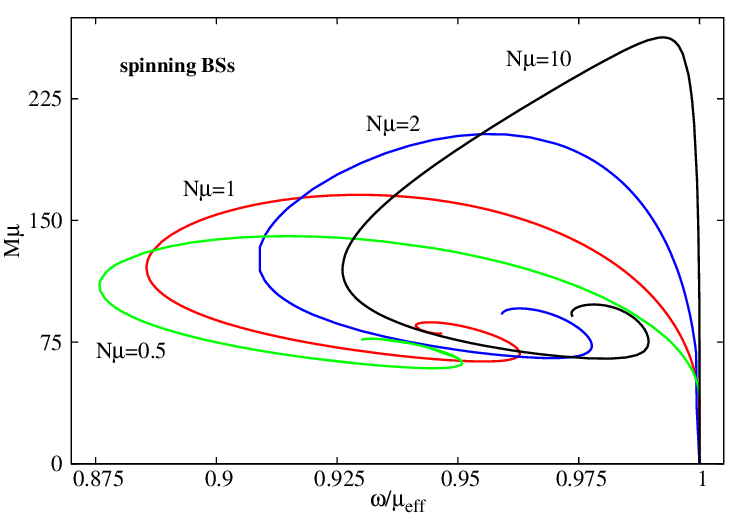}
		\end{subfigure}%
	}
	\caption{
	{\small The  frequency-mass  diagram
		is shown for static (left panel)
		and spinning (right panel) families of BSs
		with several values of the parameter $N$. 
		In all plots for solutions with 
		squashed KK asymptotics shown in this work,
		the background contribution $M_0$
		 - $cf.$ Eq. (\ref{MTvac}) -
		is subtracted from the mass $M$, such that $M=0$
    for horizonless, vacuum solutions.}
		\label{fig:2}}
\end{figure}
\begin{figure}[htbp]
	\makebox[\linewidth][c]{%
		\begin{subfigure}[b]{8cm}
			\centering
			\includegraphics[width=8cm]{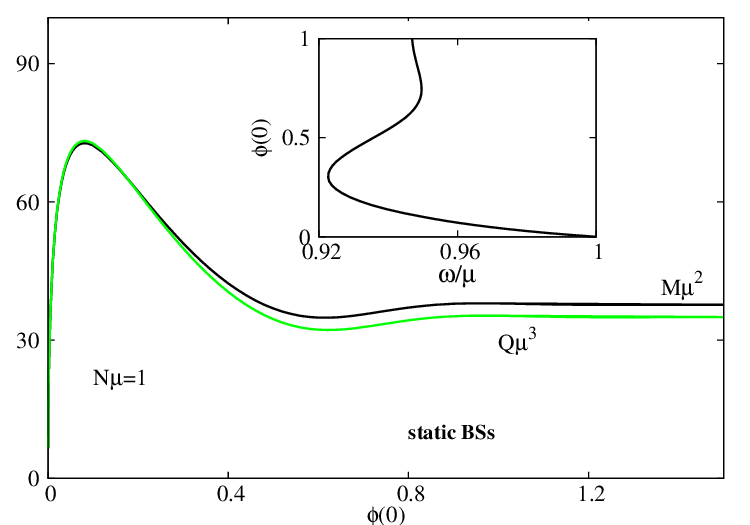}
		\end{subfigure}%
		\begin{subfigure}[b]{8cm}
			\centering
			\includegraphics[width=8cm]{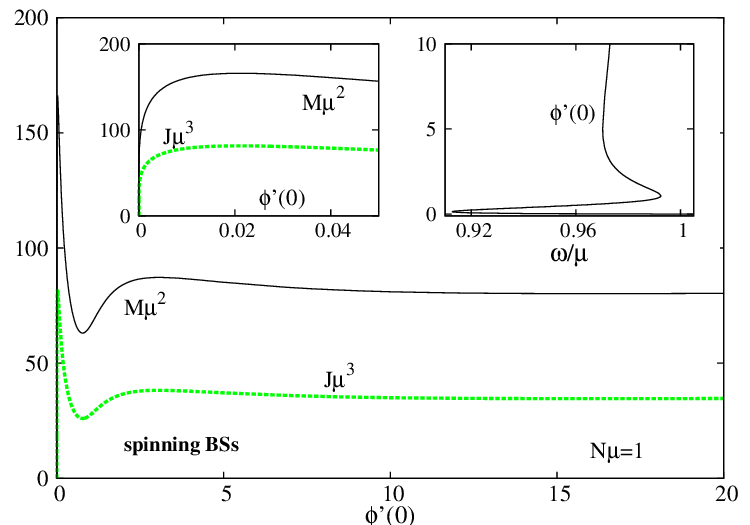}
		\end{subfigure}%
	}
	\caption{
	{\small
	The mass $M$ and Noether charge $Q$ (angular momentum $J$)
		are shown as a function of the central value of the scalar field  $\phi(0)$
		(left panel, static solutions)
		and as a function of the first derivative
		of the scalar field at the origin  $\phi'(0)$
		(right panel, spinning solutions).
		The insets shows how $\phi(0)$ (or $\phi'(0)$) 
		is related to the frequency parameter $\omega$.\label{fig:3}}
		}
\end{figure}

The static solutions are constructed for a version of 
(\ref{metricBS}) with $W=0$
and $s=0$ in the scalar field ansatz
(\ref{scalar}). 
While their far field behavior is still given by (\ref{infBS}), with $\mu_{\rm eff}=\mu$,
the scalar field does not vanish at $r=0$, with 
$\phi(r)=\phi_0 +{\cal O}(r).$

The profile of typical static and spinning BS solutions, with the same values of the input parameters 
$N,\mu$ and $\omega$, 
 are shown in Figure \ref{fig:1}.
One remarks that the metric functions $F_0,F_1,F_2$ (and 
$W$ in the rotating case)
interpolate monotonically between some constant value at the origin and zero
at infinity, without presenting a local extremum. 
   
Taking $\omega$ as a control parameter, the numerical results show that
both static and spinning  
BSs exist for a limited range of frequencies,
$\omega_{\rm min}  <\omega< \mu_{\rm eff}$,
with $\omega_{\rm min}$ depending on the value of the NUT parameter $N$. 
In the  limit  $\omega\rightarrow\mu_{\rm eff}$,
the mass, tension
and the Noether charge
 go to zero\footnote{Recall that  the  background contributions $M_0,\,\mathcal{T}_0$
are subtracted such that $M=\mathcal{T}=0$
in the absence of a scalar field,
$cf.$ the discussion in Section \ref{charges}.}.
One remarks that this is the behavior also found
in  the $d=4$ asymptotically flat case \cite{Liebling:2012fv}.

As can be seen in Figure \ref{fig:2}, 
the BS mass first increases as $\omega$ is decreased from  $\mu_{\rm eff}$
approaching a maximal value, $M_{\rm max}$, for some $\omega_0$ (with both 
$M_{\rm max}$
and
$\omega_0$
 increasing with $N$).
Then, the mass decreases, and, after some $\omega_{\rm min}$, a
backbending is observed in the $M(\omega)$-diagram. 
Further following the curve, there is an inspiralling behaviour, 
towards a limiting configuration
at the center of the spiral, which occurs for a frequency $\omega_{\rm cr}$ (which is 
also a function of $N$).
This central inspiralling behaviour appears to be generic for BS solutions
in EKG model, 
being also found in $d=4$ dimensions,
or for $d=5$ solutions with $\mathbb{M}^{1,4}$
or even AdS asymptotics.
A similar diagram is recovered for the curve $Q(\omega)$;
for static BSs,
one finds $M< \mu Q$
for a range of frequencies between $\mu$
and a critical value marked with a black dot in Figure 2 (left)
(see also the inset).
Rather unexpectedly, we have found that $M>\mu Q$
for all considered spinning configurations,
which suggests that these solutions
are unstable. 
 
Further insights on the properties of the BSs can be taken from Figure \ref{fig:3}, 
where we plot the mass $M$ and Noether charge $Q$ 
as a function of the central value of the scalar field $\phi(0)$ (static case)
 and  $\phi'(0)$ (spinning BSs). 
Again, one notices a rather similar picture 
to that found for static BSs in $d=4$
\cite{Kaup:1968zz,Ruffini:1969qy}.

\subsection{Synchronized hairy Black Holes}
\label{sec:SycnchHBHs}
 
The BH solutions are constructed for a
slightly
 more complicated metric ansatz, which 
fixes the behaviour at the horizon and at infinity, and also
can be taken as a deformation of the static vacuum BH (\ref{metric2s}), 
namely\footnote{ The spinning BS solutions can also be studied within the metric ansatz
(\ref{BHmetric})
with $r_H=0$. 
However, the numerics is more difficult as compared to the metric choice (\ref{metricBS}),
 the small-$r$ expansion 
of the scalar field starting with a $\sqrt{r}$-term.
}
\begin{align}
	\nonumber
	ds^2=-e^{2F_0(r)}\frac{\left( 1-\frac{r_H}{r} \right)^4 }{\left( 1+\frac{r_H}{r} \right)^2 }dt^2&+e^{2F_1(r)}H(r) \left( 1+\frac{r_H}{r} \right)^4  
	\bigg [
	dr^2+r^2 \left(\sigma_1^2+ \sigma_2^2 \right)
	\bigg] 
	\\	
	\label{BHmetric}
	&+e^{2F_2(r)}\frac{4N^2}{H(r)}[\sigma_3-2W(r) dt]^2\,,
\end{align}
with $F_0,F_1,F_2$ and $W$
resulting from numerics,
 and the background function  
\begin{eqnarray}
	\label{H}
	H(r)=1+\frac{2\left(\sqrt{N^2+r_H^2}-r_H\right) r }{ (r+ r_H )^2}. 
\end{eqnarray}
As with the BSs, one can write an 
 approximate form of the solutions
at the limits of the $r$-interval.
 The essential coefficients in these expansions determine most of the quantities of interest, 
either horizon quantities  ($r=r_H$) or global quantities ($r\to \infty$). 
 At the horizon, the
Killing vector\footnote{
$\xi \Psi=0$ is
the only symmetry of the full solution
(geometry and scalars)
 and is generated by
$\xi$. 
Also, the BSs are
invariant under the action of $\hat \xi=\partial_t+2\omega \partial_\psi$.
}
 $\xi=\partial_t+\Omega_H \partial_\psi$
becomes null, with $\Omega_H=2W(r_H)$ the event horizon angular velocity.
%
\begin{figure}[htbp]
	\makebox[\linewidth][c]{%
		\begin{subfigure}[b]{8cm}
			\centering
			\includegraphics[width=8cm]{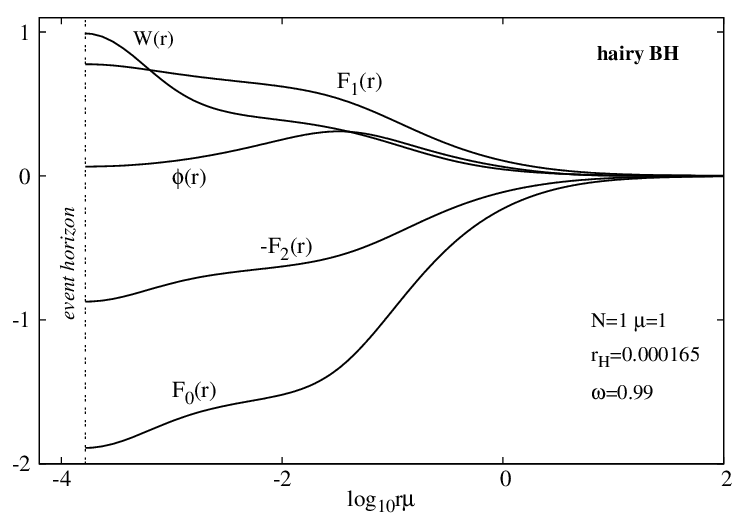}
		\end{subfigure}%
		\begin{subfigure}[b]{8cm}
			\centering
			\includegraphics[width=8cm]{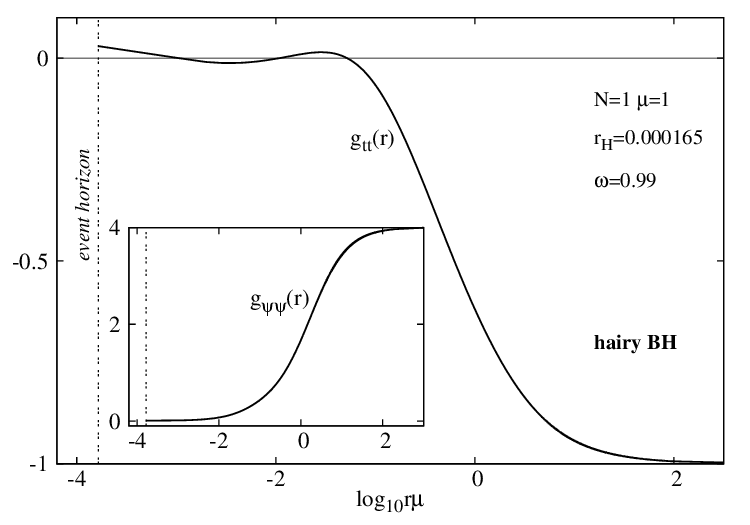}
		\end{subfigure}%
	}
	\caption{
	{\small
	(Left panel) 
		The metric functions $(F_i,W)$ 
		and the scalar field amplitude $\phi$
		of a HBH solution
		are shown as a function 
		of the radial coordinate.
		(Right panel)
		The metric functions $g_{tt}(r)$ and  $g_{\psi \psi}(r)$
		are shown for the same solution.
		One notices the existence of two ergo-regions ($g_{tt}>0$);
		also the metric function 
		$g_{\psi \psi}(r)$ interpolates smoothly between the horizon value 
		$g_{\psi \psi}(r_H)=0.0076$ and the asymptotic 
		value $g_{\psi \psi}(\infty)=4N^2$.
		}
		\label{fig:4}}
\end{figure}
 The following (formal) power series holds  there
(where $i=0,1,2$):  
\begin{eqnarray}
	\nonumber
	&&
	F_i(r)=f_{i0}^{(H)}+f_{i2}^{(H)}(r-r_H)^2+\dots, 
	\\
	&&
	\label{BHeh}
	W(r)=\frac{1}{2}\Omega_H+w_{2}^{(H)}(r-r_H)^2+\dots,~
	\phi(r)= \phi_{0}^{(H)}+ \phi_{2}^{(H)}(r-r_H)^2+\dots,~~
\end{eqnarray} 
%
with
the following relation between frequency and event horizon angular velocity 
\begin{eqnarray}
	\label{synch}
	\omega =W\big|_{r_H}=\frac{1}{2}\Omega_H ~,
\end{eqnarray} 
which is just the condition (\ref{cond1})
with $m=1/2$, as implied by the employed scalar ansatz.

 The shape of the event horizon can be read off from the induced horizon metric 
\begin{eqnarray}
	d\Sigma_H^2=8e^{2f_{10}^{(H)}}r_H^2
	\Bigg(1+\sqrt{1+\frac{N^2}{r_H^2}}\, \Bigg) 
	\left(\sigma_1^2+ \sigma_2^2 \right) 
	+\frac{8 N^2 e^{2f_{20}^{(H)}}}
	{1+\sqrt{1+\frac{N^2}{r_H^2}}} \sigma_3^2\, ,
\end{eqnarray}
which describes a squashed $S^3$ geometry. 
 The expansions at the horizon, (\ref{BHeh}), can be used to compute the event horizon's area $A_H$ and Hawking temperature $T_H$, which are given by 
\begin{eqnarray}
	A_H =  256 \sqrt{2}\pi^2 N r_H^2 e^{2f_{10}^{(H)}+f_{20}^{(H)}}
	\sqrt{1+\sqrt{1+\frac{N^2}{r_H^2}}}\,,~~
	T_H =
	\frac{1}{16 \pi r_H}
	\frac{ \sqrt{2} ~e^{f_{00}^{(H)}-f_{10}^{(H)}}  } 
	{ \sqrt{1+\sqrt{1+\frac{N^2}{r_H^2}}}}\,. \ \ \ 
\end{eqnarray}
 An approximate solution can also be constructed for large $r$, with  
\begin{eqnarray}
	&&
	\nonumber
	F_0(r)=\frac{f_{01}}{r}+\dots,~~
	F_1(r)=\frac{f_{11}}{r}+\dots,~~
	F_2(r)=-\frac{f_{11}+f_{01}}{r}+\dots,~~
	\\
	&&
	\label{BHinf}
	W(r)=\frac{w_{1}}{r}+\dots,~~~~
	\phi(r)=c_1\frac{e^{-r\sqrt{\mu^2_{\rm eff}-\omega^2}}   }{r }+\dots.
\end{eqnarray}
%
%
\begin{figure}[htbp]
	\makebox[\linewidth][c]{%
		\begin{subfigure}[b]{8cm}
			\centering
			\includegraphics[width=8cm]{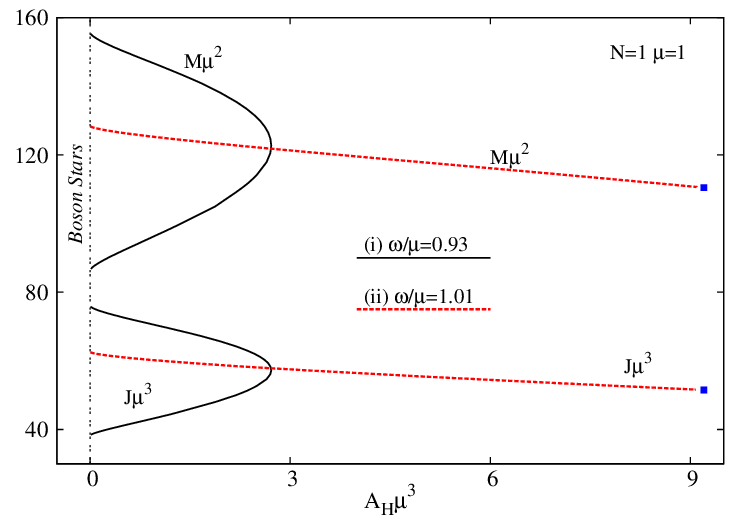}
		\end{subfigure}%
		\begin{subfigure}[b]{8cm}
			\centering
			\includegraphics[width=8cm]{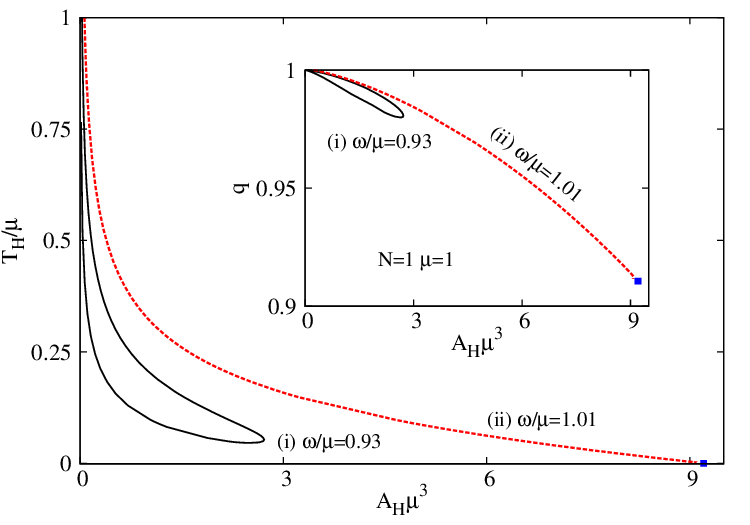}
		\end{subfigure}%
	}
	\caption{
	{\small
	The mass $M$, 
		angular momentum $J$,
		Hawking temperature $T_H$
		and hairiness parameter $q=Q/(2J)$
		are shown as a function of event horizon area $A_H$
		for two sets of solutions with field frequency $\omega$
		(or,
		equivalently, constant angular velocity $\Omega_H$ ).
		In the two cases  the HBHs interpolate between a BS
		and $(i)$  another BS or, $(ii)$ an extremal HBH.
		}
		\label{fig:5}}
\end{figure}

 With these expressions, 
the computation of the mass, tension  and angular momentum is straightforward,
with 
\begin{eqnarray}
	&&
	M=\frac{4\pi}{G_5}N
	\left(
	\sqrt{N^2+r_H^2}+3 r_H 
	+f_{11}-f_{01}
	\right),~~~J=\frac{16\pi N^3 w_1}{G_5} ,
	\\
	&&
	\nonumber
	{\cal T}=\frac{1}{G_5} 
	\left(
	\sqrt{N^2+r_H^2}+f_{11}+\frac{1}{2}f_{01}
	\right).
\end{eqnarray}
As usual in BH mechanics
(without a cosmological term), 
the temperature, horizon area and the global charges are
related through a Smarr mass formula \cite{Smarr:1972kt, Bardeen:1973gs}, 
whose general form for the squashed KK asymptotics
reads
\begin{eqnarray}
	\label{Smarr}
	M= \frac{1}{2}{\cal T} L+\frac{3}{2}T_H  \frac{A_H}{4G_5} 
	+\frac{3}{2} \Omega_H \left(J-\frac{1}{2}Q\right)+M^{(\Psi)},
\end{eqnarray}
where $L=8\pi N$ is the length of the twisted $S^1$ fiber at infinity, 
and
\begin{eqnarray}
	M^{(\Psi)}=-\frac{3}{2}\int_{\Sigma}  \sqrt{-g} d^4x  \left( T_t^t-\frac{1}{3}T_a^a \right),
\end{eqnarray}
is the mass stored in the matter field(s) outside the horizon. 

As with other BHs with synchronized hair,
to measure the 'hairiness' of the solutions,  
we introduce a $normalized$ Noether charge $q$, with  $q=0$ for a vanishing scalar field 
and $q=1$ for BSs:
\begin{eqnarray}
	q= \frac{Q}{2J}~.
\end{eqnarray}
 
The complete classification of the solutions in the space of input 
parameters 
$\{N; \omega, r_H\}$ is a considerable task
which is beyond the scope of this paper.
%
\begin{figure}
	\makebox[\linewidth][c]{%
		\begin{subfigure}[b]{8cm}
			\centering
			\includegraphics[width=8cm]{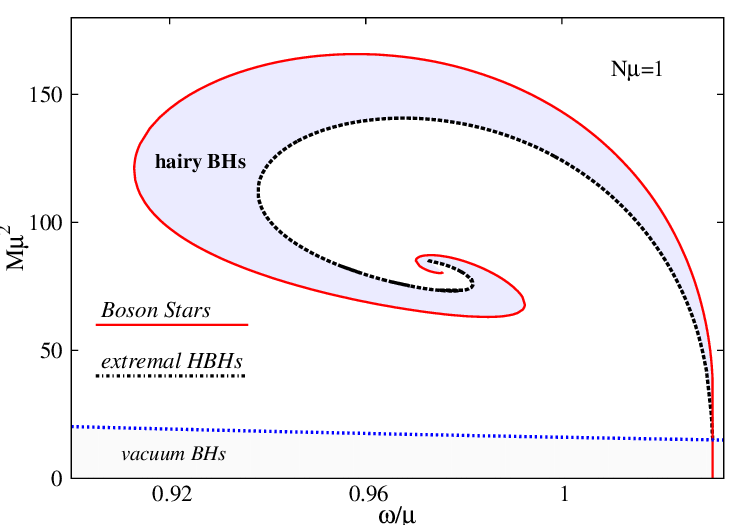}
		\end{subfigure}%
		\begin{subfigure}[b]{8cm}
			\centering
			\includegraphics[width=8cm]{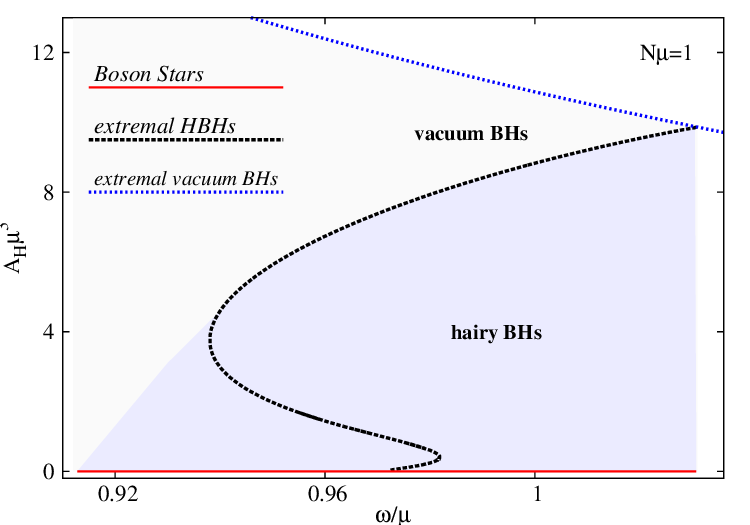}
		\end{subfigure}%
	}\\
	\caption{
	{\small
	The
		domain of existence (shaded area) 
		of HBHs
		is shown 
		in a mass $vs.$ frequency (left panel)
		 and 
		 horizon area $vs.$ frequency (right panel) plot.
	The vacuum, spinning BHs exist below the blue dotted line 
	(in which case one takes 
	$\Omega_H=2\omega$).
	}
	\label{fig:6}}
\end{figure}
%
In what follows we show results for
$N=1$,
while a very similar phase  diagram  has been 
found for $N=0.5$, 
these being the only values of $N$ for which we have attempted 
for a systematic investigation of the solutions.
However, we have also constructed  HBHs with $N=0.1,~2$ and $5$,
and the displayed picture for $N=1$ appears to be generic.
      
The profile of a BH solution which smoothly interpolates between the asymptotic expansions
(\ref{BHeh}),
(\ref{BHinf}) 
is shown in Figure \ref{fig:4}.
An interesting feature there is the existence of two distinct
ergo-regions ($i.e.$ with $g_{tt}>0$),
a feature which is found also for $d=4$
BHs with synchronized scalar hair
\cite{Herdeiro:2014jaa}.
This is not, however, the generic behaviour,
since a single ergo-region is found for a 
large part of the parameter space.
 
Given $N\neq 0$,
the domain of existence of solutions is obtained by considering sequences of solutions 
at constant $\omega=\Omega_H/2$
and varying the event horizon radius $r_H$.
As expected, a (small) BH can be added at the center of any spinning  BS with a given $\omega$; this is 
the starting point for any aforementioned sequence.
However, the end point depends on the value the frequency parameter (see Figure \ref{fig:5}).
For 
$\omega_{\rm min}<\omega<\omega_i$, 
the sequence  ends in another BS with the same frequency 
(case $(i)$ in Figure \ref{fig:5}), 
where we introduced $\omega_i$ to denote the minimum frequency possible for extremal hairy BHs,   
$i.e.$ the zero temperature configurations, 
see the black dotted curve in Figure \ref{fig:6}.
A different picture is found for 
$\omega_i<\omega<\mu_{\rm eff}$ (case $(ii)$), 
the sequences ending on  extremal BHs
with a nonvanishing horizon area and hairiness parameter $q$.
  
In Figure \ref{fig:6} (left panel), 
we exhibit the domain of existence of the 
HBHs  (shaded region), in a $M(\omega)$ diagram, based on around two thousands of solution points, 
 a similar diagram being found for $J(\omega)$.
The picture possesses some similarities 
with the one found for spinning BHs with scalar hair
in $d=4$ 
\cite{Herdeiro:2014goa,Herdeiro:2015gia}.
As with that case,  the BS curve (red solid line in Figure \ref{fig:6})
forms a boundary of the domain; in particular the BSs set  the maximal 
value of the BHs' mass.
\begin{figure}[h]
	\makebox[\linewidth][c]{%
		\begin{subfigure}[b]{8cm}
			\centering
			\includegraphics[width=8cm]{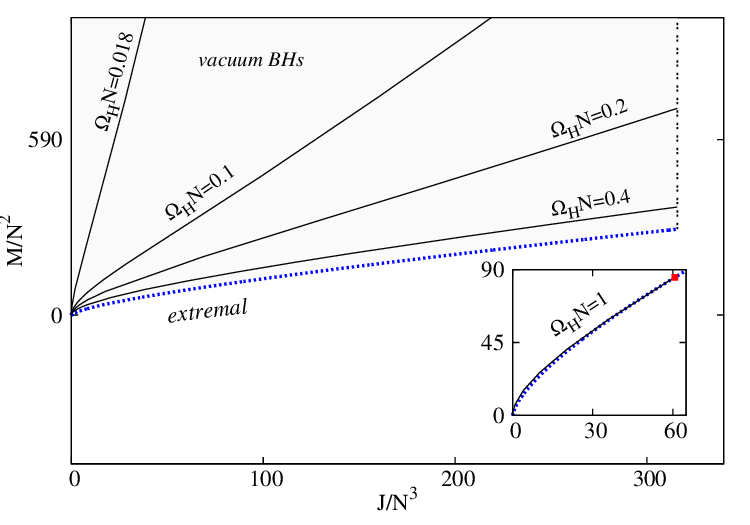}
		\end{subfigure}%
		\begin{subfigure}[b]{8cm}
			\centering
			\includegraphics[width=8cm]{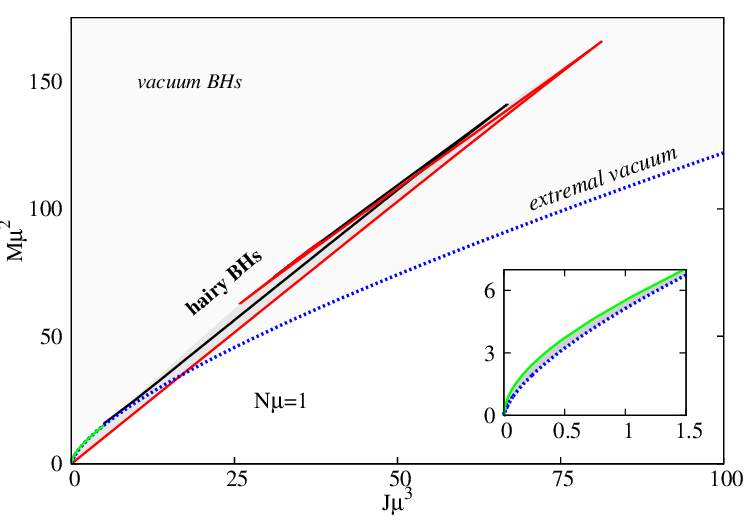}
		\end{subfigure}%
	}\\
	\caption{
	{\small
	The mass-angular momentum diagram is shown for 
		vacuum BHs (left panel) and for HBHs 
		(right panel).
		}
		\label{fig:7}}
\end{figure}
There is also a  curve of \textit{extremal} 
HBs (black line),
which appears to inspiral towards a
central value, where, we conjecture, it meets the endpoint of the BS spiral.
In $d=4$, the extremal BHs curve ends 
 in a particular vacuum Kerr solution,
where it joins the {\it existence line} -- 
a particular set of Kerr BHs allowing for scalar clouds \cite{Herdeiro:2014goa,Herdeiro:2015gia}.
However, in the absence  of an  existence line for $d=5$ - no scalar clouds on a vacuum BH backgound -, 
the extremal HBH curve continues all
the way to the maximal frequency, $\omega=\mu_{\rm eff}$.

Now consider 
the horizon area $vs.$ 
frequency diagram, see Figure \ref{fig:6} (right panel).
Differently from the $d=4$ case 
\cite{Herdeiro:2015gia},
one notices the existence there of 
a vertical line segment with $\omega=\mu_{\rm eff}$ and nonzero horizon area. 
That is, these solutions exist
for a given range of $r_H \geqslant 0$;
there
the scalar field spreads and tends to zero as $\omega$
increases towards $\mu_{\rm eff}$,
and also the values of the Noether charge $Q$ and of 
the mass stored in the field $M^{(\Psi)}$ both
tend to zero.
At the same time, the geometry does not trivialize,
becoming that of a vacuum (spinning)
KK BH, with the input parameters\footnote{The range of event horizon
radius here is $0\leqslant r_H\leqslant r_H^{\mathrm{ext}}$, the limits corresponding to 
the GPS soliton and the extremal vacuum BH, respectively.}
$\Omega_H=2\mu_{\rm eff}$ and $r_H$
 (and nonzero global charges
 $M,{\cal T}$ and $J$).  

Further insight can be found in Figure \ref{fig:7} (right panel),
where we plot the domain of existence of HBHs in the
($J,M$)-plane 
Again, despite the absence of an existence line,
the overall picture is rather similar to that found in 
\cite{Herdeiro:2015gia} 
for the $d=4$ HBHs counterparts. 
Observe, however, the existence in Figure \ref{fig:7} (also in the inset) 
of a green line,
which corresponds
to the limiting solutions with $\omega=\mu_{\rm eff}$;  
this line starts at vacuum and ends  in an extremal vacuum BH with (for $N\mu=1$)
 $M=15.0334$ and $J=4.9348$.

\section{ Boson stars and Black Holes with squashed Kaluza-Klein asymptotics $vs.$ 	
solutions with $\mathbb{M}^{1,4}$ and $\mathbb{M}^{1,3}\times S^1$ 	asymptotics}
\label{sec:Asympt}

 One of the main goals of this work 
is to identify the effects of considering squashed KK asymptotics on the 
properties of BSs and hairy BHs solutions of the EKG system. 
Here we recall that, as discussed in Section \ref{sec:Background}, 
the GPS soliton - whose asymptotics provide the background
of our solutions - 
can be seen as a vacuum state interpolating between the $\mathbb{M}^{1,4}$ and the 	(standard) 	
$\mathbb{M}^{1,3}\times S^1$ vacua, which are approached in the limit of an infinite $N$ and vanishing $N$, respectively.
 As such, it is of interest to contrast the picture found above for EKG solutions with 
$N\neq 0$,
 with that for BSs and BHs solutions of the same model (\ref{action}), 
which, however, approach at infinity a background given by (\ref{metricM5}) or (\ref{metricM4S1}).
 This comparison will be the main subject of this Section.

\subsection{Solutions with $\mathbb{M}^{1,4}$ asymptotics}

The case of EKG solutions with  $\mathbb{M}^{1,4}$ asymptotics
is better understood, being the subject of 
several studies.
Starting with   static, spherically symmetric solutions,
we consider 
the scalar ansatz (\ref{scalarS})
and the following metric form
with two functions $U(r)$ and $\delta(r)$
\begin{eqnarray}
	\label{metricM5d}
	ds = - e^{-2\delta(r)} U(r) dt^2+
	\frac{dr^2}{U(r)}+\frac{r^2}{4}
	\left[ 
	d\theta^2+\sin^2 \theta d\varphi^2+(d\psi+\cos \theta d\varphi)^2
	\right] ,
\end{eqnarray}
the line-element 
(\ref{metricM5}) being approached asymptotically.
The horizonless solitonic solutions have
been discussed in  
\cite{Hartmann:2010pm},
describing spherically symmetric BSs,
and share all basic properties of the spinning stars discussed below.

By adapting a general theorem put forward in 
\cite{Pena:1997cy},
one can show that,
as with the $d=4$ case and  $\mathbb{M}^{1,3}$ asymptotics,
there are no static, spherically symmetric BHs
with scalar hair
(here we assume the existence of an horizon with 
$U(r_H)=0$ and $\delta(r_H)$ finite).
The starting point is the conservation of the energy-momentum tensor of
the scalar field 
\begin{eqnarray}
	\nabla_a T^a_b=0~,
\end{eqnarray}
which, for $b\equiv r$  and the considered ansatz (note that we take $s=0$ in 
(\ref{scalar})), 
results in 
\begin{eqnarray}
	\label{rel1}
	%
	e^{ \delta}( e^{-\delta}T_r^r)'=-\frac{3}{r}T_r^r+\frac{1}{2}\frac{dg_{ab}}{dr}T^{ab},
\end{eqnarray}
with
\begin{eqnarray}
	\label{rel1n}
 T_r^r = U\phi'^2+\left(\frac{e^{2\delta}\omega^2 }{U}-\mu^2\right)\phi^2,
\end{eqnarray}
and
\begin{eqnarray}
	\frac{1}{2}\frac{dg_{ab}}{dr}T^{ab}&=&  \\
\nonumber
	-\frac{3U\phi'^2}{r}
	&+&\left(\frac{e^{2\delta}\omega^2}{U}-\mu^2\right)\frac{3\phi^2}{r}
	+
	\left[
	U\phi'^2+\left(\mu^2+\frac{e^{2\delta}\omega^2}{U}\right)\phi^2
	\right]\delta'
	-\left(\phi'^2+\frac{e^{2\delta}\omega^2 \phi^2}{U^2}\right)U'.
\end{eqnarray}
However, $U'$ and $\delta'$
can be eliminated from the above relation
 by using a suitable combination of the Einstein equations\footnote{
	The Einstein equations implies 
\begin{eqnarray}
\label{Eeqd5}
	 \delta'+\frac{4}{3} r\left(\phi'^2+\frac{e^{2\delta}\omega^2 \phi^2}{U^2}\right)=0,~~~
~~
 [r^2(1-U)]'=\frac{4}{3} r^2\left[U\phi'^2+\left(\mu^2+\frac{e^{2\delta}\omega^2 }{U}\right)\phi^2\right] .
	\end{eqnarray}
},
which results in the following form of  Eq. (\ref{rel1})
\begin{eqnarray}
	\label{rel11}
	e^{ \delta}( e^{-\delta}T_r^r)'= -\frac{2e^{2\delta}w^2(1-U)\phi^2}{r U^2}
	-\frac{2(1+2U)\phi'^2}{r} .
\end{eqnarray} 
One notices that,
 since the   $U$-equation 
in (\ref{Eeqd5})
implies
$U<1$,
  the $r.h.s.$ of the above relation is a strictly negative quantity.
Therefore $e^{-\delta}T_r^r$
is a strictly decreasing function.  
Moreover, the equation (\ref{rel11})
implies the following relation
(where we use the fact that the scalar field decay exponentially at infinity
and thus  $T^r_r\left(\infty\right)=0$):
\begin{equation}
\label{sup1s}
	T^r_r(r_H)=e^\delta(r_H) \intop_{r_H}^\infty \mathrm{d} \bar r\;
	e^{-\delta}\left(\frac{2e^{2\delta}w^2(1-U)\phi^2}{ \bar r U^2}+\frac{2(1+2U)\phi'^2}{ \bar r} \right) \geqslant 0~.
\end{equation}  
To analyze the near horizon  limit
of Eq. (\ref{rel11}),
 one introduces a proper radial distance  $x$, 
which is regular at the horizon, with
  $dx=\frac{dr}{\sqrt{U}}$. In terms of this coordinate, the Eq. (\ref{rel11}) becomes: 
\begin{eqnarray}
	\label{rel11s}
	e^{ \delta}\frac{d (e^{-\delta}T_r^r)}{dx}= 
	-\frac{2e^{2\delta}w^2(1-U)\phi^2}{r U^{3/2}}
	-\frac{2\sqrt{U}}{r} (1+2U)\phi'^2~. 
\end{eqnarray} 
 For regular solutions, the $r.h.s.$ of this equation should remain finite as
$U\rightarrow0$.  
It follows that the quantity $\frac{2e^{2\delta}w^2 \phi^2}{r U^{3/2}}$
must remain finite as $r\to r_H$.
Thus the term $\frac{e^{2\delta}\omega^2 }{U}\phi^2$ will vanish in the same limit, 
which, from (\ref{rel1n}), implies $T_r^r(r_H)\leqslant 0$. 
However, this would contradicts the Eq. (\ref{sup1s})
unless $T_r^r(r_H)=0$. 
Moreover, since the integrand of the $r.h.s.$ in (\ref{sup1s})
has a negative sign,
it follows that $\phi=0$,
$i.e.$ the absence of scalar hair.

\medskip

\begin{figure}
	\makebox[\linewidth][c]{%
		\begin{subfigure}[b]{8cm}
			\centering
			\includegraphics[width=8cm]{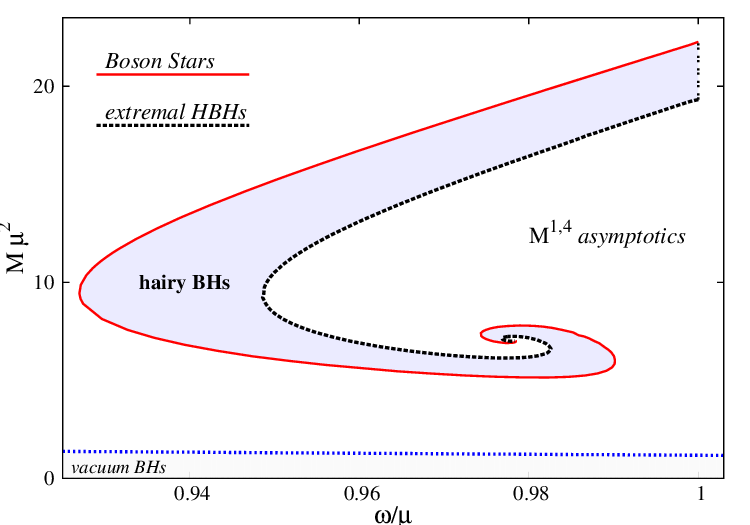}
		\end{subfigure}%
		\begin{subfigure}[b]{8cm}
			\centering
			\includegraphics[width=8cm]{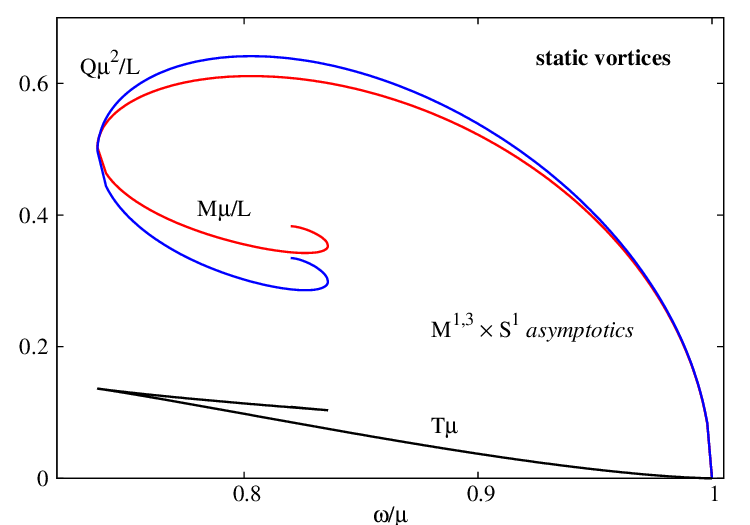}
		\end{subfigure}%
	}\\
	\caption{
	{\small
	{\it Left}: The (frequency-mass) domain of existence of hairy BHs with $\mathbb{M}^{1,4}$  
	(adapted from~\cite{Brihaye:2014nba}). 
	{\it Right}: The mass, Noether charge and tension are shown as a function of frequency for EKG static
	vortices with $\mathbb{M}^{1,3} \times S^1$ asymptotics.
	}
	\label{fig:8}}
\end{figure}

Turning to the case of the scalar field
with dependence on the coordinates on the three-sphere, 
$i.e.$
the
ansatz
(\ref{scalar})
with $s=1$, the 
corresponding spinning BSs have
been discussed in~\cite{Hartmann:2010pm}. 
Their most striking
property  is that they
 do not trivialize as the maximal frequency is reached, $\omega \to \mu$
(note that this also holds  for static BSs).
While in this limit the scalar field spreads and tends to zero point-wise
 and the geometry becomes 
arbitrarily close to that of  $\mathbb{M}^{1,4}$, 
the BS mass (and Noether charge/angular momentum)
remains {\it finite} and {\it nonzero} in that limit, see 
the red curve in 
Fig. \ref{fig:8} (left panel).
This implies the existence of a gap 
between the $\omega \to \mu$ 
limiting configurations and the  $\phi=0$ (vacuum) $\mathbb{M}^{1,4}$ ground state. 
This is very different from the case of a $\mathbb{M}^{1,3}$ background,
where all BS charges vanish as $\omega \to \mu$,
both in the static and in the spinning cases\footnote{Moreover, 
	a similar behaviour is found for 
	the $d=4$ static non-spherically symmetric BSs 
	reported in~\cite{Herdeiro:2020kvf}, which can be axially symmetric chains  
	or even configurations with no spatial isometries.}. 
An analytical argument   which
helps to understand this different behavior was presented in~\cite{Hartmann:2010pm},
which we shall briefly review here. 
This relies on the special scaling properties of the EKG system,
which are dimension  dependent.
Basically, as $\omega \to \mu$,
the  radial coordinate and the scalar field  scale  as
$r=\tilde r/\xi$, 
$\phi=\xi^2 \tilde \phi$
where 
$\mu^2=\omega^2 +\xi^2 \hat w_c^2$,
with
$\xi$ a small parameter and $\hat w_c$ a constant.
Then, in $d=4$
the integral $\int_0^{\infty}  dr~ r^2 \phi^2$
(which determines the Noether charge) vanishes
as $\xi \to 0$,
while the $d=5$  corresponding expression
$\int_0^{\infty}  dr~ r^3 \phi^2$
remains finite and nonzero.  
The same reasoning explains the different behavior of
the scalar field mass-integral   for $d=4,~5.$

This argument also helps to partially understand the
behavior we  have found above for the BSs
with squashed KK asymptotics.
Since the size of $S^3$ in the generic ansatz (\ref{metric})
becomes proportional in the far field with $r^2$ only,
the solutions are effectively four dimensional
and thus the Noether charge integral (or the integral for the mass stored
in the field)
vanishes as $\omega \to \mu_{\rm eff}$.

Differently from the spherically symmetric case, 
the  
scalar field ansatz  (\ref{scalar}) with $s=1$
allows for hairy spinning BH solutions 
\cite{Brihaye:2014nba},
which are found for the same ansatz employed in this work,
and also obey the 
synchronization condition
(\ref{synch}).
As with the BH solutions in Section \ref{sec:TheSolutions},
the hair of the asymptotically $\mathbb{M}^{1,4}$ is intrinsically non-linear,
without the existence of
scalar clouds on a vacuum Myers-Perry BH background \cite{Myers:1986un}
($i.e.$ of an existence line). 
Additionally, and naturally, the asymptotically $\mathbb{M}^{1,4}$  BH solutions 
inherit from the solitonic limit
a gap for the mass, angular momentum and Noether charge - Fig. \ref{fig:8} (left panel).

\subsection{The  $\mathbb{M}^{1,3}\times S^1$ case: EKG vortices and no hairy Black Strings}
\label{sec:M4timesS1}

To our best knowledge, 
the case of EKG solutions
with  $\mathbb{M}^{1,3}\times S^1$ asymptotics (see Eq.
(\ref{metricM4S1})),
has not yet been considered 
in the literature. 
Such solutions, if they exist, describe EKG 
vortices 
and Black Strings.

To study them,
we consider the scalar ansatz (\ref{scalarS}) together with
  the following  line element 
\begin{eqnarray}
	\label{metricKK}
	ds^2=e^{-a \psi (r)} 
	\left(-\e^{-2 \delta(r)} U(r) dt^2+
	\frac{dr^2}{U(r)}+r^2 d\Omega_2^2
	\right)
	+e^{2a \psi (r)} dz^2~,
\end{eqnarray}
which makes contact
with the $d=4$ picture discussed in the next Section, 
with $a=2/\sqrt{3}$
and an arbitrary periodicity $L$ for the $z$-coordinate.
The metric functions 
$\psi (r)$,
$\delta(r)$,
$U(r)$,
and
the scalar amplitude
$\phi(r)$
are solutions of the equations
\begin{eqnarray}
\nonumber
	&&
	(e^{-\delta}r^2 U \psi')'+2a e^{-a\psi-\delta}\mu^2 r^2\phi^2=0\;,~~
	\delta'+r \psi'^2+2 r \left(\phi'^2+\frac{e^{2\delta}\omega^2\phi^2}{U^2}\right)=0,
	\\
	\label{KK-BS1}
	&&
	[r(1-U)]'-2r^2 \left(U \phi'^2+(\mu^2 e^{-a\psi})+\frac{e^{2\delta}\omega^2}{U}\right)\phi^2
	-r^2 U\psi'^2=0,
	\\
\nonumber
	&&
	(e^{-\delta}r^2 U\phi')'+r^2\left(\frac{e^{-\delta}\omega^2}{U}-e^{-a\psi-\delta}\mu^2\right)\psi=0~.
\end{eqnarray}
The vortices
have no horizon, 
the size of the $S^2$ sector in (\ref{metricKK})
shrinking to zero as $r\to 0$,
while the size of the $z$-circle remains finite,
with 
$\psi(r)=\psi_0+{\cal O}(r^2)$,
$U(r)=1+{\cal O}(r^2)$,
$\delta(r)=\delta_0+{\cal O}(r^2)$
and
$\phi(r)=\phi_0+{\cal O}(r^2)$
close to $r=0$ (where 
$\psi_0$, 
$\delta_0$,
$\phi_0$ are parameters fixed by numerics).
The behaviour for large-$r$ is
\begin{eqnarray}
	\label{KK-BS-inf}
	\psi=\frac{\psi_1}{r}+\dots,~
	U=1+\frac{h_1}{r}+\dots,~
	\delta=\frac{\psi_1^2}{2r^2}+\dots,~
	\phi=c_1 \frac{e^{-r\sqrt{\mu^2-\omega^2}}}{r}+\dots,~
\end{eqnarray}
with the free parameters $\psi_1$, 
$h_1$ and 
$c_1$.

The vortices posses a nonvanishing mass, tension\footnote{
	The mass and tension are computed following~\cite{Harmark:2003dg,Harmark:2004ch}.
	However, a similar result is found within the counterterm approach,
	with the same boundary term (\ref{countertermaction}).}
and Noether charge, with
\begin{eqnarray}
	\label{KK-BS-charges}
	M=-\frac{h_1 L}{2G_5},~~
	{\cal T}=-\frac{h_1+3 a \psi_1}{4 G_5},~~
	Q= 8\pi\omega L \int_0^\infty dr\frac{r^2 e^{\delta} \phi^2}{U}~.
\end{eqnarray}
The frequency-mass diagram of these solutions is shown in Figure \ref{fig:8} (right panel).
The picture there strongly resembles that for $d=4$
spherically symmetric BSs in the (pure) EKG model \cite{Liebling:2012fv}.
This can be understood by noticing that, when performing a KK reduction
$w.r.t.$ the $z$-direction,
these EKG vortices become $d=4$
BSs in a EKG-dilaton model - see Section 5.

\medskip 

As expected, no Black Strings with complex scalar hair exist in this case. 
This can be shown following the same P\~ena-Sudarsky-type argument
\cite{Pena:1997cy}
 as in the previous subsection. 
It is straightforward to show that the conservation of the stress-energy tensor 
together with the Einstein equations implies the following relation
\begin{eqnarray}
	\label{rel11n}
	e^{ \delta+a\psi}( e^{-\delta- a \psi}T_r^r)'=  
	-\frac{e^{2\delta+a \psi} \omega^2(1-U)\phi^2}{r U^2}
	-\frac{e^{a \psi}(1+3U)\phi'^2}{r} 
 -\mu^2 r \phi^2 \psi'^2
	+a \mu^2 \phi^2 \psi' .~~~~{~~~~}
\end{eqnarray} 
Since    the $\psi$-equation in  (\ref{KK-BS1})
implies that $\psi'<0$,
we conclude that $T_r^r$  is a strictly decreasing function, with $T_r^r(r_H) \geqslant 0$.
However, by using similar arguments as employed above to rule out 
the existence of (spherically symmetric) BHs with   $\mathbb{M}^{1,4}$  asymptotics,
that is, by rewriting the eq. (\ref{rel11n}
in terms of 
the proper radial distance $dx=\frac{dr}{\sqrt{U}}$),
one finds that  $T_r^r(r_H)\leqslant 0$.  
This leads to a contradiction,
 and we conclude that the scalar field necessarily vanishes.

The case of solutions with the $s=1$ scalar field ansatz 
(\ref{scalar})  
and $\mathbb{M}^{1,3}\times S^1$  asymptotics
is unclear and we leave it for future studies.

\section{The  Kaluza-Klein reduction and the four dimensional picture
	}
\label{sec:4Dpic}

The solutions with squashed KK asymptotics
in Section \ref{sec:TheSolutions}
and also the above discussed vortices can be considered from a $d=4$ perspective, upon KK reduction on a circle.
Following  \cite{Gross:1983hb},
let us consider a generic  KK metric ansatz
\begin{align}
	\label{metric5}
	ds_5^2&= e^{-a \psi (x)} ds_4^2+e^{2a \psi (x)} (dz+2A_i(x) dx^i)^2\,,\\
	&\quad	{\rm with}~~ds_4^2=g_{ij}^{(4)}(x) dx^i dx^j~~{\rm and}~~a=\frac{2}{\sqrt{3}},~{~~}
\end{align} 
where in this section $x^i$ denote  $d=4$ coordinates, with time $t$ being one of them, 
and $z$ is the (compact) fifth-dimension, which has some periodicity $L$.

For the scalar field (which can be a multiplet), 
one only assumes that it has a specific  
$z$-dependence, which disappears at the level of 
energy-momentum tensor and
equations of motion,
a single mode being excited,
\begin{equation} 
	\label{scalarN} 
	\Psi = \Phi(x)  e^{i k z  } ,
\end{equation}  
where the function $\Phi $ can be complex, and
 $k=2 \pi m/L$ ($m=0,\pm 1, \pm 2,\dots$).

As such, the $d=5$ EKG system
admits an equivalent four dimensional description,
the function $g_{zz} $
determining the dilaton $\psi$, while the 
metric components
$g_{iz}$ resulting in a U(1) field, with the field strength tensor
$F_{ij}=\partial_i A_j-\partial_j A_i$. 
That is, after integrating over the $z-$coordinate and
dropping  a boundary term,
the resulting four dimensional  action reads
\begin{eqnarray}
	\label{action4} 
	&&
	\mathcal{S}_4 = \frac{1}{4 \pi G_4}\int_\mathcal{M}  d^4x \sqrt{-g^{(4)}}\bigg[  
	\frac{1}{4}R^{(4)} 
	-\frac{1}{4} e^{3 a \psi} F_{ij} F^{ij} 
	-\frac{1}{2} \partial_i\psi \partial^i \psi
	\\
	\nonumber
	&&
	{~~~~~~~~~~~~~~~~~}
	-\frac{1}{2} g^{ij(4)}
	\left(
	D_{i}\Phi^\dagger  D_{j}\Phi +  
	D_{j}\Phi^\dagger  D_{i}\Phi 
	\right) 
	-U (|\Phi|,\psi \big) 
	\bigg] ~,
\end{eqnarray}
with
$G_4=G_5/L$.
Observe that the $d=4$ scalar field $\Phi$ is $gauged$ 
$w.r.t.$ the U(1) field $A_i$,
with the  gauged 
derivative
\begin{eqnarray}
	D_{j}\Phi =(\partial_j -i q_s A_j  )\Phi,~~ 
\end{eqnarray} 
the gauge coupling constant being $q_s=2k$
and a  potential\footnote{
For a  dilaton $\psi$ which vanishes aymptotically,
	the $d=4$ scalar $\Phi$
	possesses an \textit{effective} mass
	$	\mu_{\rm eff}^2= \mu^2+k^2.$
}
\begin{eqnarray} 
\label{Un}
	U(|\Phi|,\psi )= \mu^2 |\Phi|^2 e^{-a \psi} +k^2  e^{-3a \psi} |\Phi|^2,
\end{eqnarray} 
which depends on both $\Phi$ and $\psi$. This EdMgs model reduces to the standard KK Einstein-dilaton-Maxwell (EdM) model for $\Phi=0$. For vanishing gauge potential, $A_i=0$, it reduces to an Einstein-dilaton-Klein-Gordon (EdKG) model.

\subsection{ Static, dyonic Black Holes with gauged scalar hair}
\label{sec:4Dpicd}

Starting with the vacuum case $(\Psi=0)$,
and the squashed KK asymptotics,
let us remark that,
since $g_{zz}$ shrinks to zero as $r\to 0$, the dilaton diverges there and 
the five-dimensional GPS soliton is singular from a four dimensional perspective
\cite{Gross:1983hb,Sorkin:1983ns}.
However, the situation is different with BHs; for example, the 
spinning solution in Section \ref{app:AppVacBH} results in 
spherically symmetric dyonic BHs in a specific Einstein-Maxwell-dilaton model
discussed $e.g.$ in~\cite{Gibbons:1985ac}.

Turning now to the EKG solution in Section \ref{sec:TheSolutions},
the situation with the BSs is similar to that of (vacuum) KK monopoles, being singular at $r=0$ 
in the $d=4$ picture.
The spinning HBHs, on the other hand, result  in a 
family of $d=4$ solutions 
of the model (\ref{action4})
which describe
static spherically symmetric BHs with resonant gauged scalar hair
and a dyonic U(1) field.
All properties of these solutions follow  from those of the $d=5$
EKG BHs.
To make explicit this correspondence,   
we have found useful to  consider an (equivalent) version of
the $d=5$ line element
with the following form of the last term in  Eq. (\ref{BHmetric})
\begin{eqnarray}
\label{newS3}
	e^{2F_2(r)}\frac{1}{H(r)}(dz+2N\cos \theta  d\varphi-4N W(r) dt)^2,~~
	{\rm with}~~~z=2N\psi~,
\end{eqnarray}
such that the dilaton $\psi(r)= {F_3(r)}/{a}$ 
vanishes asymptotically\footnote{Alternatively, one can work with the initial form
 in  Eq. (\ref{BHmetric}),
 which implies a non-vanishing
	dilaton in the far field, and then consider a 
	rescaling of the $d=4$  line element.}.
The $d=4$ gauge field describes a 
dyon, with
$A=V(r) dt+ Q_m \cos\theta d\varphi$,
the $N$-parameter becoming in the $d=4$ perspective  the magnetic charge, 
 $Q_m= N$,
while the $W(r)$-function associated with rotation 
determines the electric potential, 
$V(r)=-2N W(r)$.
The $d=4$ BH metric reads
\begin{eqnarray}  
	\label{BHmetric4}
	&&
	ds_4^2=-S_0(r)dt^2+  S_1(r)
	\big[
	dr^2+r^2 (d\theta^2+\sin^2 \theta d\varphi^2)
	\big]
	,
	~~{~~}
	\\
	&&
	{\rm where}~~
	S_0(r)=e^{2F_0(r)+a F_3(r)}
	\frac{\left( 1-\frac{r_H}{r} \right)^4 }
	{\left( 1+\frac{r_H}{r} \right)^2},~~
	S_1(r)=e^{2F_1(r)+a F_3(r)}H(r) \left( 1+\frac{r_H}{r} \right)^4,
	\nonumber
\end{eqnarray}
and $H(r)$ is given by Eq. (\ref{H}).
The expression of the $d=4$ scalar field reads\footnote{
	Properties of this scalar ansatz 
	(including regularity)
	are discussed in a more 
	general context in~\cite{Gervalle:2022npx,Gervalle:2022vxs}.}
\begin{equation}
	\label{scalar4}
	\nonumber
	\Phi = \phi(r) 
	\left( 
	\begin{array}{c} 
		~\sin\frac{\theta}{2}~e^{-i \frac{\vphi}{2}} 
		\\    
		\cos\frac{\theta}{2}~e^{i \frac{\vphi}{2}} 
	\end{array} 
	\right) 
	e^{-i\omega t} .
\end{equation} 

The gauged coupling constant is fixed
by the $N$-parameters, with
$q_s=1/(2N)$ (and $k=1/(4N)$.
Then one can easily see that $d=5$
\textit{synchronization condition}
(\ref{cond1})
translates into the 
$d=4$
\textit{resonance condition}
(\ref{cond2}), that is\footnote{This result holds as well when considering a
form of the metric 
 (\ref{BHmetric})
with the coordinate $\psi$ replaced by $z$, $cf.$ Eq. (\ref{newS3}).
Although in this case $\Omega_H=4NW({r_H})$,
the synchronization condition $\omega= W({r_H})$ still holds,
since
 the $z$-dependence in (the new form of) scalar field ansatz implies $m=1/(4N)$. 
}
\begin{eqnarray}
	\omega= m \Omega_H=W\big|_{r_H}=-\frac{1}{2N}V\big|_{r_H}=q_s {\cal V},
\end{eqnarray} 
with ${\cal V}=V(\infty)-V(r_H)$
the electrostatic chemical potential\footnote{
Since $ W(\infty)=0$, the $d=4$ BH solutions are found by fixing
a (residual) gauge freedom via $V(\infty)=0$,
while~\cite{Herdeiro:2020xmb} 
uses $V(r_H)=0$.
The physical results are,
of course, independent of the gauge choice.}.
 
There is a simple map between the all quantities of interest
of the  $d=5$ BHs
and those of the $d=4$ solutions, with $e.g.$
\begin{eqnarray}
\nonumber
	L G_4 M^{(d=4)}=G_5 M^{(d=5)},~L G_4 Q_e^{(d=4)}=2 G_5 J^{(d=5)},~
	A_H^{(d=4)}=\frac{1}{L}A_H^{(d=5)},~T_H^{(d=4)}=T_H^{(d=5)},
	~~{~~~}
\end{eqnarray} 
where $Q_e$ is the electric charge.

Differently from the other 
BHs with
\textit{resonant hair}
discussed
in the literature
\cite{Hong:2019mcj,Herdeiro:2020xmb,Hong:2020miv,Brihaye:2022afz},
these $d=4$ solutions exist  without 
self-interaction terms in $\Psi$, in the scalar potential (\ref{Un}).
Finally, we remark that the absence of an existence line 
for the $d=5$ vacuum spinning BHs
corresponds to the absence of charged scalar bound states
on the background of a dyonic BH in a KK Einstein-dilaton-Maxwell (EdM) model, although
this result should not perhaps  be a surprise, given the similar findings
in \cite{Hod:2012wmy,Hod:2013nn}
 for  the Reissner-Nordstr\"om BH case.

\subsection{ Other cases}
\label{sec:4Dpicds}
The case of EKG vortices with $\mathbb{M}^{1,3}\times S^1$
asymptotics in Section \ref{sec:M4timesS1}
is simpler, since the direct KK reduction
leads to an (ungauged) EdKG model, $i.e.$ with $A_i=0$ in (\ref{action4}),
while the $d=4$ metric form and dilaton are read directly from  (\ref{metricKK}). 
We remark that, for the employed $s=0$ ansatz  (\ref{scalarS}), (\ref{scalar}),
 it is more natural to interpret the results as for a model with a single scalar field,
which has the same expression (and also field mass $\mu$) in both $d=4,~5$.

$d=4$ solutions with a gauged scalar field can, nonetheless,   
be generated by using the $d=5$ EKG  vortices
as seeds.
The basic procedure is well known in the literature - 
however, without also considering a complex scalar field - 
and works as follows.
Starting with any EKG vortex,
we perform a boost in the $(t,z)$-plane,
with 
\begin{equation}
	\begin{cases}
		t=\cosh \alpha~ T-\sinh \alpha~ Z\\
		z=\cosh \alpha ~Z-\sinh \alpha ~T\\
	\end{cases}\,,~~{\rm where}~~\alpha\in\mathbb{R}\,.
\end{equation}
Then a KK reduction $w.r.t.$
the direction $Z$ 
results in the following
solution of the $d=4$ model
(\ref{action4})
\begin{eqnarray}
	\label{metric4-new}
	ds_4^2=-\frac{ e^{-2 \delta(r)} N(r)}{\sqrt{S(r)} } dT^2+\sqrt{S(r)} \left(\frac{dr^2}{N(r)}+r^2 d\Omega_2^2 \right)
	,~~
	A=V(r)dt,
\end{eqnarray}
where
\begin{eqnarray} 
	\nonumber 
	S(r)=\cosh^2\alpha-  ~e^{-\delta(r)-3a \psi(r)}N(r)\sinh^2\alpha,~~
	V(r)= \left( e^{-2 \delta(r)-3a \psi(r)}N(r)-1\right) \frac{\sinh \alpha \cosh \alpha}{2S(r)}~,
\end{eqnarray}
while the $d=4$ dilaton field is
$\psi(r)=\psi_i(r) +\frac{\log S(r)}{2a},$
with
$\psi_i(r)$
the function which enters the seed $d=5$ 
solution, denoted by $\psi$ in the metric (\ref{metricKK}).
The $d=4$ scalar field $\Phi$ is
\begin{eqnarray} 
	\Phi=\phi(r) e^{-i  \tilde \omega T },
	~~{\rm with}~~ \tilde \omega=\omega \cosh \alpha ,
\end{eqnarray}
while  the gauge coupling constant 
is
$g_s=2\omega \sinh \alpha $.

These solutions describe  static, electrically charged,
spherically symmetric  
gauged BSs, which are
a generalisation of the usual (uncharged) EKG BSs~\cite{Jetzer:1989av,Jetzer:1992tog,Pugliese:2013gsa}, 
with an extra-dilaton field, $cf.$ eq. (\ref{action4}). 
Again, it is straightforward to derive the map
between the
quantities of interest of the $d=5$ and $d=4$
configurations. 

\medskip

Finally, we mention that the  same setup
can be used to construct a 
generalization of the known $d=4$ spherically symmetric (ungauged)
BSs by including the effects of a background $U(1)$ magnetic field.
In this case, one starts again with the $d=5$ 
EKG vortices in Section \ref{sec:M4timesS1};
however, the resulting $d=4$
configurations have axial symmetry only,
and describe (unguged) BSs
which approach asymptotically a dilatonic Melvin Universe background.
The way to introduce a $d=4$ magnetic field in a KK setup involves twisting the 
$z$-direction \cite{Dowker:1994up,Dowker:1995gb},
that is by taking $\varphi \to \varphi+B_0 z$
in the metric (\ref{metricKK}),
with $B_0$ an arbitrary real constant, and
reidentifying points appropriately.
Upon reduction, the resulting  $d=4$ solutions have a
line element
\begin{eqnarray}
	\label{metric4-Melvin} 
	ds_4^2=\sqrt{\Lambda(r,\theta)} 
	\left(
	\frac{dr^2}{N(r)}+r^2 d\theta^2 -N(r)e^{-2\delta(r)}dt^2 
	\right)
	+\frac{r^2 \sin^2 \theta d\varphi^2}{\sqrt{\Lambda(r,\theta)}},
\end{eqnarray}
with
$\Lambda(r,\theta)=1+ e^{-3a\psi(r)}B_0^2 r^2 \sin^2 \theta$.
The  U(1)-potential and the $d=4$ dilaton are  
\begin{eqnarray}
	A=\frac{e^{-3a\psi(r)}B_0  r^2 \sin^2 \theta}{2 \Lambda(r,\theta)}d\varphi,~~
	\psi(r,\theta)=\psi_i(r)+\frac{1}{2a} \log  \Lambda(r,\theta),
\end{eqnarray}
while the scalar $\Phi$ has the same expression as $\Psi$ in five dimensions,
and remains $ungauged$. 

\section{Conclusions and remarks}
\label{sec:Conclusions}

The study of solitons and BH solutions in more than $d= 4$
dimensions is a subject of long
standing interest in General Relativity,
the case of a KK  theory
with only one (compact) extra-dimension 
providing the simplest model. 
Although the original  proposal
in
\cite{kaluza,Klein:1926tv}
does not result in a realistic theory of Nature, it still continues to provide
insight
into more sophisticated theories, such as  
supergravity  and string/M-theory.

In the context of this work, we were mainly interested
in KK solutions of the $d=5$  Einstein equations
with a squashed sphere at infinity,
the simplest case being the vacuum soliton found 
by Gross and Perry 
\cite{Gross:1983hb}
and Sorkin
\cite{Sorkin:1983ns}
(GPS).
As discussed by various authors,  an horizon can be added also for these asymptotics,
which results in BHs with a squashed horizon of
$S^3$ topology
 \cite{Chodos:1980df,Dobiasch:1981vh,Pollard:1982gj,Gibbons:1985ac,Wang:2006nw}.

The main purpose of this paper was to extend the  
solutions in~\cite{Gross:1983hb,Sorkin:1983ns,Chodos:1980df,Dobiasch:1981vh,Pollard:1982gj,Gibbons:1985ac,Wang:2006nw} 
by including a scalar field doublet, with a special ansatz, originally proposed in~\cite{Hartmann:2010pm}, 
in the action of the model;
both solitons (BSs) and BHs were considered.

\medskip

Our main results can be summarized as follows:
\begin{itemize} 	
\item
We have provided evidence for the
existence of BHs with scalar hair, 
with the same far field  squashed KK asymptotics as the
 GPS soliton.
These solutions provide further evidence for the universality
of the \textit{hair synchronization}
mechanism.
They satisfy the same specific condition
between the scalar field frequency
and event horizon velocity
known to hold for a variety of other BHs with (complex) scalar hair,
see $e.g.$
\cite{Herdeiro:2014goa}-\cite{Herdeiro:2017oyt}.
	Moreover, similar to other cases,
	the BHs do not trivialize in the limit of
	a vanishing horizon area, and become
	BS solutions with squashed KK asymptotics.

	\item
	The basic properties of the considered BSs and BHs
	are a combination between those of the known
	$d=4$  and $d=5$  solutions
	with Minkowski spacetime asymptotics.
	For example, as with the $d=4$ case \cite{Liebling:2012fv}, 
	the global charges of the BSs 
	vanish as the maximal frequency is approached.
	On the other hand, for BHs, there is no {\it existence line},
	$i.e.$ of scalar clouds on a vacuum, spinning BH background, 
	a feature shared with the $d=5$ solutions in Ref. \cite{Brihaye:2014nba}.

\item
	As a new feature induced by the 
	squashed KK asymptotics, 
	the scalar field possesses 
	(in the spinning case) an 
	effective mass  $\mu_{\rm eff}^2=\mu^2+\frac{1}{16 N^2}$,
	where the second term is a geometric contribution -
	$N$ is related to the size of the twisted
	$S^1$ fiber at infinity.
	The bound state condition for the scalar field
	frequency is  $\omega\leqslant \mu_{\rm eff}$.  
 
	\item
 These  $d=5$ solutions of the EKG equations 
	possesses an equivalent 
	$d=4$ description.
	While, as with the vacuum GPS case \cite{Gross:1983hb},
	the solitons corresponds to $d=4$ singular configurations,
	the KK reduction of the BHs result
in static spherically symmetric dyonic BHs with gauged scalar hair,
in a specific EdMgs model.

\end{itemize}

\begin{figure}
	\makebox[\linewidth][c]{%
		\begin{subfigure}[b]{14cm}
			\centering
			\includegraphics[width=14cm]{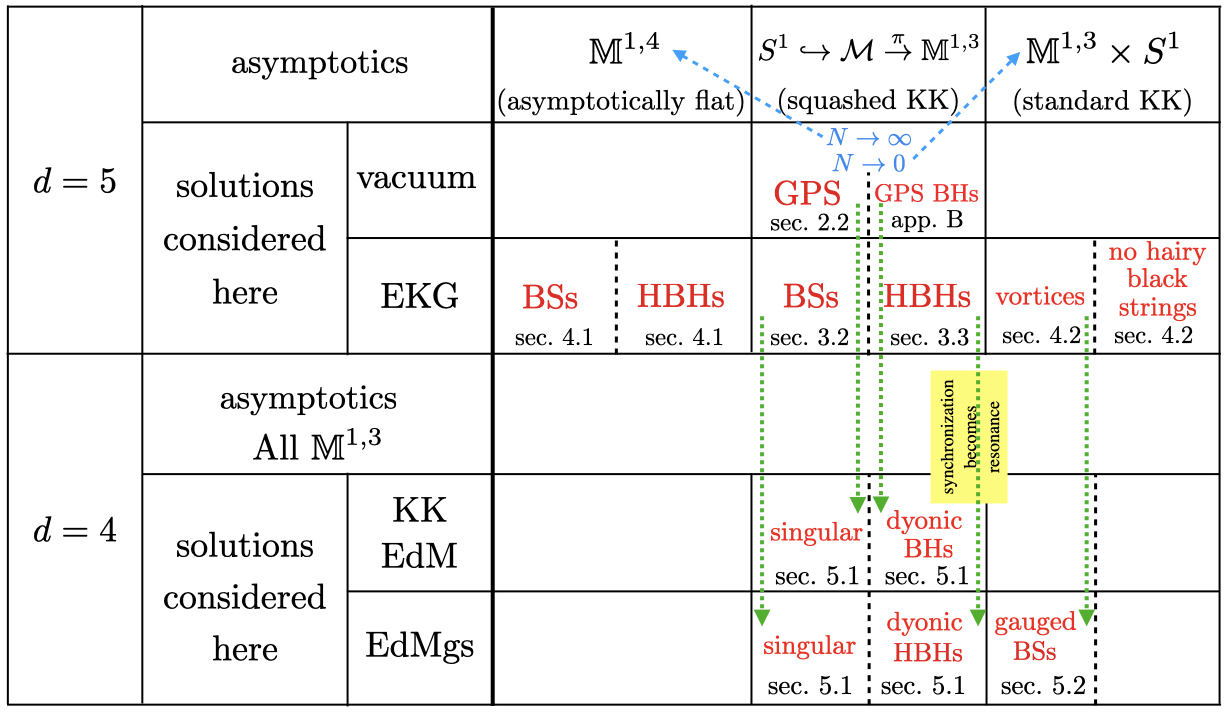}
		\end{subfigure}%
	}\\
	\caption{
	{\small
	Overview of the solutions used or constructed in this paper and their inter-relations. The vertical green dashed arrows represent KK reduction. The sections where the solutions are discussed in this paper are also given.
	}
	\label{fig:9}}
\end{figure}

As a byproduct of this study, we have also  investigated 
  EKG solutions with  standard $\mathbb{M}^{1,3}\times S^1$  asymptotics
	and established first the absence of
	static Black Strings with scalar hair.
However,  horizonless solutions do exist, corresponding to
EKG vortices.
After boosting and considering a KK reduction,
these configurations result in 
spherically symmetric, charged BSs,
generalizing for an extra-dilaton field
the known gauged BSs~\cite{Jetzer:1989av,Jetzer:1992tog,Pugliese:2013gsa}. Figure~\ref{fig:9} provides an overview of the solutions studied in this paper and their relations.

\medskip

As possible avenues for future research,
we mention first the more systematic investigation 
 of the  
$d=5$ squashed BHs solutions,  such as the study of geodesic motion, their lensing properties or their thermodynamics,
together with a detailed study of
the resulting $d=4$ configurations.
It would be interesting to investigate  
 solutions with the same squashed KK asymptotics,
but which 
rotate also in $\varphi$-direction, for the coordinates in the metric ansatz (\ref{metric}).
This would result in EKG generalization of the vacuum 
BHs discussed in~\cite{Rasheed:1995zv,Matos:1996km,Larsen:1999pp};
their study, however, requires solving a set of partial differential equations.
Moreover, the KK reduction of these HBHs would result in
a generalization of the $d=4$
Kerr-Newman BHs with scalar hair
studied in \cite{Delgado:2016jxq},
with an extra-dilaton field in the action and also with a nonzero magnetic charge.

Finally, we mention the case of 
KK dipolar asymptotics instead of  monopolar, as in this work.
In the vacuum case, such a configuration
has been considered in~\cite{Gross:1983hb},
being again of the form
(\ref{metric-gen}),
with $ds_4^2$ there
corresponding to the Kerr instanton metric.
We anticipate the existence of similar configurations in an  EKG model with a single
complex
scalar field. 
Differently from the vacuum case, which necessarily possess a 'bolt' - 
$i.e.$ a minimal nonzero size of the $S^2$ part in  $ds_4^2$ metric -,
the  EKG system would allow also also for 'nutty' solitons -
that is, with the  $S^3$ part in the  $ds_4^2$ metric shrinking
to zero at $r=0$, as with the BSs in this work. 
We hope to return elsewhere with a study of these aspects.

\acknowledgments  
C.H., J.N. and E.R. 
would like to express their gratitude to Juan Carlos Degollado for being an exceptional host
during a visit at at the Instituto de Ciencias Físicas, UNAM - Campus de Morelos, Mexico,
where a part of this project has been developed. 
This work is supported by the Center for
Research and Development in Mathematics and Applications (CIDMA) through the Portuguese Foundation for Science and Technology (FCT – Fundação para a Ciência e a Tecnologia), references UIDB/04106/2020 and UIDP/04106/2020. The authors acknowledge support from the projects PTDC/FIS-AST/3041/2020, as well as CERN/FIS-PAR/0024/2021 and 2022.04560.PTDC. This work has further been supported by the European Union’s Horizon 2020 research and innovation (RISE) programme H2020-MSCA-RISE-2017 Grant No. FunFiCO-777740 and by the European Horizon Europe staff exchange (SE) programme HORIZON-MSCA-2021-SE-01 Grant No. NewFunFiCO-101086251.  J. N. is supported by the FCT grant 2021.06539.BD.

\appendix
\section{The Einstein and energy-momentum tensors }
\label{app:Eqs}

To get an idea about the equations solved in practice, we display here the
expression of the 
 non-vanishing components of the Einstein tensor $E_a^b=R_a^b-\frac{1}{2}R\delta_a^b$ 
and 
of energy-momentum tensor $T_a^b$  for the generic metric ansatz (\ref{metric}) 
and the scalar ansatz (\ref{scalarS}), (\ref{scalar}).
For the Einstein tensor, one finds
\begin{align}
	\nonumber
	E^t_t&=\frac{1}{ {\cal F}_{2}}\left(\frac{{\cal F}_{3}}{4{\cal F}_{2}}-1\right)
	+\frac{1}{{\cal F}_{1}}\left(\frac{{\cal F}_{2}'{}^{2}}{4{\cal F}_{2}^{2}}
	+\frac{{\cal F}_{3}'{}^{2}}{4{\cal F}_{3}^{2}}+\frac{{\cal F}_{1}'{\cal F}_{2}'}{{\cal F}_{1}F_{2}}-\frac{{\cal F}_{3}'{\cal F}_{2}'}{{\cal F}_{2}F_{3}}+\frac{{\cal F}_{1}'{\cal F}_{3}'}{2{\cal F}_{1}{\cal F}_{3}}\right)
	\\
	\nonumber
	&-\frac{1}{{\cal F}_{1}}\left(\frac{2{\cal F}_{3}WW''}{{\cal F}_{0}}
	+\frac{{\cal F}_{2}''}{{\cal F}_{2}}+\frac{{\cal F}_{3}''}{2{\cal F}_{3}}
	+\frac{{\cal F}_{3}W'^{2}}{{\cal F}_{0}}\right)
	-\frac{{\cal F}_{3}WW'}{{\cal F}_{0}{\cal F}_{1}}\left(\frac{{\cal F}_{0}'}{{\cal F}_{0}}+\frac{{\cal F}_{1}'}{{\cal F}_{1}}-\frac{2{\cal F}_{2}'}{{\cal F}_{2}}-\frac{3{\cal F}_{3}'}{{\cal F}_{3}}\right)\,,
	\\
	\label{Einstein-tensor}
	E_r^r&=\frac{{\cal F}_3 W'^2}{{\cal F}_0 {\cal F}_1}
	+\frac{1}{2{\cal F}_1}\left(\frac{{\cal F}_2'^2}{2 {\cal F}_2^2}
	+\frac{{\cal F}_0' {\cal F}_2'}{{\cal F}_0 {\cal F}_2}
	+\frac{{\cal F}_3' {\cal F}_2'}{{\cal F}_2 {\cal F}_3}
	+\frac{{\cal F}_0' {\cal F}_3'}{2 {\cal F}_0 {\cal F}_3}\right)
	+\frac{1}{{\cal F}_2}\left(\frac{F_3}{4
		{\cal F}_2}-1\right)\, ,
	\\
	\nonumber
	E_\theta^\theta&=E_\varphi^\varphi=
	(\frac{{\cal F}_0''}{{\cal F}_0}+\frac{{\cal F}_2''}{{\cal F}_2}
	+\frac{{\cal F}_3''}{{\cal F}_3})\frac{1}{2{\cal F}_1}
	-(
	\frac{{\cal F}_0'^2}{{\cal F}_0^2}+\frac{{\cal F}_2'^2}{{\cal F}_2^2}+\frac{{\cal F}_3'^2}{{\cal F}_3^2}
	)\frac{1}{4{\cal F}_1}
	-\frac{{\cal F}_3W'^2}{{\cal F}_0 {\cal F}_1}
	-\frac{{\cal F}_3}{4{\cal F}_2^2}
	\\
	\nonumber
	&\qquad\qquad
	+(
	-\frac{{\cal F}_0'{\cal F}_1'}{{\cal F}_0{\cal F}_1}
	+\frac{{\cal F}_0'{\cal F}_2'}{{\cal F}_0{\cal F}_2}
	+ \frac{{\cal F}_0'{\cal F}_3'}{{\cal F}_0{\cal F}_3}
	-\frac{{\cal F}_1'{\cal F}_2'}{{\cal F}_1{\cal F}_2}    
	-\frac{{\cal F}_1'{\cal F}_3'}{{\cal F}_1{\cal F}_3}    
	+          -\frac{{\cal F}_2'{\cal F}_3'}{{\cal F}_2 {\cal F}_3}        
	)\frac{1}{4{\cal F}_1},
	 \\
		\nonumber
	E^\psi_\psi&=\frac{1}{{\cal F}_{1}}\left(-\frac{2{\cal F}_{3}WW''}{{\cal F}_{0}}
	+\frac{{\cal F}_{0}''}{2{\cal F}_{0}}+\frac{{\cal F}_{2}''}{{\cal F}_{2}}\right)
	-\frac{1}{{\cal F}_{1}}\left(\frac{3{\cal F}_{3}W'^{2}}{{\cal F}_{0}}
	+\frac{{\cal F}_{0}'{}^{2}}{4{\cal F}_{0}^{2}}+\frac{{\cal F}_{2}'{}^{2}}{4{\cal F}_{2}^{2}}\right)
 \\ 
	\nonumber
	&\qquad+\frac{{\cal F}_{3}WW'}{{\cal F}_{0}{\cal F}_{1}}
	\left(\frac{{\cal F}_{0}'}{{\cal F}_{0}}+\frac{{\cal F}_{1}'}{{\cal F}_{1}}
	-\frac{2{\cal F}_{2}'}{{\cal F}_{2}}-\frac{3{\cal F}_{3}'}{{\cal F}_{3}}\right)
	-\frac{1}{2{\cal F}_{1}}\left(\frac{{\cal F}_{0}'{\cal F}_{1}'}{2{\cal F}_{0}{\cal F}_{1}}
	+\frac{{\cal F}_{2}'{\cal F}_{1}'}{{\cal F}_{1}{\cal F}_{2}}
	-\frac{{\cal F}_{0}'{\cal F}_{2}'}{{\cal F}_{0}{\cal F}_{2}}\right)	
	\\
	&\nonumber
	\qquad+\frac{1}{{\cal F}_{2}}\left(\frac{3{\cal F}_{3}}{4{\cal F}_{2}}-1\right),
	\\
	\nonumber
			E^t_\psi &=\frac{1}{\cos \theta} E^t_\varphi =\frac{W'}{{\cal F}_{1}}\left(\frac{{\cal F}_{3}{\cal F}_{0}'}{2{\cal F}_{0}^{2}}
	+\frac{{\cal F}_{3}{\cal F}_{1}'}{2{\cal F}_{0}{\cal F}_{1}}
	-\frac{{\cal F}_{3}{\cal F}_{2}'}{{\cal F}_{0}{\cal F}_{2}}
	-\frac{3{\cal F}_{3}'}{2{\cal F}_{0}}\right)-\frac{{\cal F}_{3}W''}{{\cal F}_{0}{\cal F}_{1}},\\
        \nonumber
        E_\varphi^\psi&= \cos\theta\left[\frac{1}{2{\cal F}_{1}}\left(
	-\frac{4{\cal F}_{3}WW''}{{\cal F}_{0}}+\frac{{\cal F}_{2}''}{F_{2}}
	-\frac{{\cal F}_{3}''}{{\cal F}_{3}}\right)+\frac{1}{{\cal F}_{2}}
	\left(\frac{{\cal F}_{3}}{{\cal F}_{2}}-1\right)\right.
	\\
	&
	\nonumber
	\qquad\qquad\qquad+\frac{{\cal F}_{3}W'}{{\cal F}_{1}}\left(\frac{W}{{\cal F}_{0}}
	\left(\frac{{\cal F}_{0}'}{{\cal F}_{0}}+
	\frac{{\cal F}_{1}'}{{\cal F}_{1}}-\frac{2{\cal F}_{2}'}{{\cal F}_{2}}
	-\frac{3{\cal F}_{3}'}{{\cal F}_{3}}\right)-\frac{2W'}{{\cal F}_{0}}\right)
	\\
	&
	\nonumber
	\left.\qquad\qquad\qquad+\frac{1}{4{\cal F}_{1}}\left(\frac{{\cal F}_{3}'{}^{2}}
	{{\cal F}_{3}^{2}}-\frac{{\cal F}_{0}'{\cal F}_{3}'}{{\cal F}_{0}{\cal F}_{3}}
	+\frac{{\cal F}_{1}'{\cal F}_{3}'}{{\cal F}_{1}{\cal F}_{3}}
	-\frac{{\cal F}_{2}'{\cal F}_{3}'}{{\cal F}_{2}{\cal F}_{3}}
	+\frac{{\cal F}_{0}'{\cal F}_{2}'}{{\cal F}_{0}{\cal F}_{2}}
	-\frac{{\cal F}_{1}'{\cal F}_{2}'}{{\cal F}_{1}{\cal F}_{2}}\right)\right],
\end{align}
while
the non-vanishing components of the energy momentum tensor 
are
\begin{eqnarray}
	&&
	\nonumber
	T^t_t=-\mu^2\phi ^2-\frac{\omega^2-s W^2}{{\cal F}_0}\phi ^2 
	-\frac{s}{4}  \left(\frac{1}{{\cal F}_3}+\frac{2}{{\cal F}_2}\right)\phi ^2-\frac{\phi
		'^2}{F_1},
	\\
	&&
	\nonumber
	T_r^r=\frac{\phi'^2}{{\cal F}_1}-s \frac{1}{2} (\frac{1}{{\cal F}_2}+\frac{1}{2{\cal F}_3})\phi^2
	+\frac{(\omega-s W)^2}{{\cal F}_0}\phi^2-\mu^2 \phi^2,
	\\
	&&
	\nonumber
	T_\theta^\theta=T_\varphi^\varphi=-\frac{\phi'^2}{{\cal F}_1}-s \frac{1}{4{\cal F}_3}\phi^2
	+\frac{(\omega-s W)^2}{{\cal F}_0}\phi^2-\mu^2 \phi^2,
	\\
	&&
	\nonumber
	T_\psi^\psi=\frac{ \left(\omega
		^2-p W^2\right)}{{\cal F}_0}\phi^2-\mu^2\phi^2+\frac{s}{4} \phi ^2 
	\left(\frac{1}{{\cal F}_3}-\frac{2}{{\cal F}_2}\right)-\frac{\phi
		'^2}{F_1},
	\\
	&&
	T^t_\psi=s\frac{ (\omega -W)}{{\cal F}_0}\phi ^2,
	~~
	T^t_\varphi=T^t_\psi\cos\theta,
	\\
	&&
	\nonumber
	T^\psi_\varphi=s\left(\frac{2 W (\omega
		-W)}{{\cal F}_0}-\frac{1}{2 {\cal F}_2}+\frac{1}{2{\cal F}_3}\right)\cos \theta \; \phi ^2,
\end{eqnarray} 
where $s=0,1$ -- $cf.$ the scalar ansatz  (\ref{scalarS}), (\ref{scalar}).

In the numerics, we choose a metric gauge with ${\cal F}_2= \lambda {\cal F}_1 r^2$ 
with $\lambda= 1/4,1$ for solitons and BHs, respectively. 
Then $E_\theta^\theta$ is a linear combination of $E_\psi^\psi$ and $E_\varphi^\psi$
(and also for the corresponding $T_a^b$)
 and we are left with five Einstein equations for four metric
functions. 
However, the $(r,r)$-Einstein equation is treated as a constraint, being satisfied once the remaining equations are zero.
Therefore we conclude that the considered ansatz is consistent.

For completeness, we include here also the general equation satisfied by the scalar amplitude: 
\begin{align}
	\label{seq}
	\phi''&+
	\frac{1}{2}
	\left(
	\frac{{\cal F}_0'}{{\cal F}_0}
	-\frac{{\cal F}_1'}{{\cal F}_1}
	+\frac{2{\cal F}_2'}{{\cal F}_2} 
	+\frac{{\cal F}_3'}{{\cal F}_3}
	\right)\phi'
	\\
	\nonumber
	&+\left[
	\frac{(\omega-s W)^2}{{\cal F}_0}-\mu^2
	-\frac{s}{2}\left(\frac{1}{{\cal F}_2}+\frac{1}{2{\cal F}_3}\right)
	\right]{\cal F}_1 \phi =0.
\end{align} 

\section{The vacuum Black Hole solution  }
\label{app:AppVacBH}


\subsection{The  static  Black Hole}

The GPS solution allows for  BH  generalizations
 \cite{Chodos:1980df,Dobiasch:1981vh,Pollard:1982gj,Gibbons:1985ac,Wang:2006nw}.
The static case has a particularly simple form, with 
\begin{equation}
	\label{metric2si}
	ds =-\bigg(1-\frac{r_h}{r}\bigg) dt^2+\left(1+\frac{2 \bar N }{r}\right)
	\left[ \frac{dr^2}{1-\frac{r_h}{r}}
	+  r^2  (d\theta^2+\sin^2 \theta d\varphi^2)
	\right]
	+\frac{4N^2}{1+\frac{2  \bar N }{r}}  (d\psi+\cos \theta d\varphi)^2\nonumber
\end{equation} 
with
$\bar N =\sqrt{N^2+\frac{1}{16}r_h^2} -\frac{1}{4} r_h.$
%
The coordinate transformation (with $r_h=4r_H)$
\begin{eqnarray}
	r \to r\left(1+\frac{r_H}{r}\right)^2 \ ,
\end{eqnarray} 
leads to the following
 equivalent form of (\ref{metric2si}) in isotropic coordinates, which was used as the background for the non-extremal HBH solutions 
reported in Section \ref{sec:SycnchHBHs}:
\begin{equation}
	\label{metric2s}
	ds^2=- \frac{\left( 1-\frac{r_H}{r} \right)^4 }{\left( 1+\frac{r_H}{r} \right)^2 }dt^2+  \left( 1+\frac{r_H}{r} \right)^4  
	\bigg[
	dr^2+r^2  (d\theta^2+\sin^2 \theta d\varphi^2)
	\bigg]
	+ \frac{4N^2}{H(r)}
	(d\psi+\cos \theta d\varphi)^2  \ , 
	~~{~~~}
\end{equation}
with the  function  $H(r)$ given by (\ref{H}) and $r_H>0$   the event horizon radius. 
Note that the Schwarzschild Black String is recovered as $N\to 0$, while $r_H=0$ 
results in the GPS solution with the form (\ref{metric2}) of $ds_4^2$ part of the metric.
The corresponding expressions
of various quantities of interest
results straightforwardly from those of the spinning BHs
displayed below.

\subsection{The rotating Black Hole}
\label{sec:AppRotBH}
\subsubsection{The general case}

The rotating generalization of the static line element (\ref{metric2s})  
has been derived in~\cite{Dobiasch:1981vh,Wang:2006nw}. 
It can be written in the generic form (\ref{metric}), where we have found 
useful to use a form of the metric functions with
\begin{eqnarray}
	&&
	\nonumber
	{\cal F}_0=\frac{(1+u^2 (1-y))}{y}
	\frac{ 1-\frac{(1-y) P_1(r)}{P_2^2(r)}  }
	{ 1+\frac{u^2(1-y) P_1^2(r)}{P_2^2(r)} }\,,
	\qquad
	{\cal F}_1= \frac{P_1(r)}{P_2(r)}\frac{y}{1-(1-y) \frac{P_1(r)}{P_2^2(r)}},~~
	\\
	&&
	\label{metric-rotation}
	{\cal F}_2= r^2 P_1(r)P_2(r)\,,\quad 
	{\cal F}_3=\frac{r_0^2\left(1+u^2 (1-y)\frac{P_1^2(r)}{P_2^2(r)} \right) P_2(r)}
		{(1+u^2)^2 y P_1(r)}\,,
	\\
	\nonumber
	&&
	W(r)=\frac{(1+u^2)^{3/2}(y-1)u}{2r_0 \sqrt{1+(1-y)u^2}}
	\frac{1-\frac{P_2^2(r)}{P_1^2(r)}}{(y-1)u^2-\frac{P_2^2(r)}{P_1^2(r)}},
\end{eqnarray} 
where
\begin{eqnarray}
	P_1(r)=1+\frac{r_0}{r},~~
	P_2(r)=1+\frac{r_0}{r} \frac{u^2}{1+u^2},~~
\end{eqnarray} 
 $\{r_0,u,y\}$ being three parameters. 
The static limit is recovered for $u=0$, resulting in the metric form (\ref{metric2si}) 
(with $r_h=r_0(1-y)/y$ and $N=r_0/(2\sqrt{y})$).

\medskip

Returning to the spinning case, we notice
first the absence of a rotating generalization of the horizonless (vacuum) GPS soliton\footnote{
Note, however, the existence of such spinning solitons in a model 
with a $U(1)$ field \cite{Tomizawa:2008rh}.}. 
To better understand the BH properties, 
we express the $y$-parameter in terms of the horizon radius $r_H$, with
\begin{eqnarray}
	y=\frac{r_0(r_H-(r_0+r_H)u^4) }{r_H(r_0+r_H)(1+u^2)^2}~,
\end{eqnarray} 
and define
\begin{eqnarray}
	r_0=s r_H~.
\end{eqnarray} 
As such, the  input parameters become $u,s$ and $r_H>0$,
with $0\leqslant u \leqslant 1$ and $1-(1+s)u^2 \geqslant 0$.

One can easily verify that metric has the right asymptotics, 
with $W\to 0$ as $r\to \infty$,
while the following relation holds
 between the horizon radius and the NUT-parameter:
\begin{eqnarray}
	%
	r_H=2N \left(1+u^2\right)  
	\sqrt{\frac{1-(1+s)u^4}{s(1+s+(3+2s)u^2+3(1+s)u^4+(1+s)^2 u^6)}}~.
\end{eqnarray}
The computation  of various quantities which enter the thermodynamic description of this
rotating BH solution 
is straightforward,  with
\begin{align}
	\nonumber
	M=&\frac{4\pi N^2}{G_5}
	\frac{
		\left(2+s +2 (1+s) u^2\right)
		\left(1+2 u^2+(1+s)u^4\right)		
	}
	{
		\left(1+s+(3+2s)u^2+3(1+s) u^4+(1+s)^2 u^6\right)^{3/2}
		\sqrt{s(1-(1+s)u^4)}\
	}\\
	\nonumber
	&\times\left(1+s+(1+2s)u^2-(1+s)u^4-(1+s)^2u^6\right)\,,
\\
	\nonumber 
	J=&\frac{16 \pi N^3}{G_5}
	\sqrt{1+s}u(1+(1+s)u^2)^2
	\\
	\nonumber
	&\times\left(
	\frac{1+u^2}{1+s+(3+2s)u^2+3(1+s) u^4+(1+s)^2u^6)}
	\right)^{3/2},
\end{align}
\begin{align}
	\nonumber
	{\cal T}=&
	\frac{N}{2G_5}
	\frac{
		\left(1+s+(1+2s)u^2 -(1+s)u^4-(1+s)^2 u^6\right)	
	}
	{
		\sqrt{s(1-(1+s)u^4)}
		(1+s+(3+2s)u^2+3(1+s)u^4+(1+s)^2 u^4)^{3/2}
	}
	\\
	\nonumber
	&\times\left(s^2u^4(u^2-1)+(1+u^2)^3+2s(1+u^2)(1+u^4) \right)\,,
\\
	&\nonumber
	\\
	\label{Kerr-quant}
	A_H=&128 \pi^2 N^3 (1+\frac{1}{s})(1+(1+s)u^2)^2 
	(1-(1+s)u^4)
	\\
	\nonumber&
	\times
	\left(
	\frac{1+u^2}{1+s+(3+2s)u^2+3(1+s) u^4+(1+s)^2u^6)}
	\right)^{3/2}\,,
	\\
	&\nonumber
	\\
	\nonumber
	T_H=&\frac{1}{8\pi N}
	\frac{
		s(1-(1+s)u^2)
		(1+s+(3+2s)u^2+3(1+s)u^4+(1+s)^2u^6)
	}
	{
		(1+s)\sqrt{s(1+u^2)}
		(1+(1+s)u^2)(1-(1+s)u^4)^{3/2}
	}\,,
	\\
	&\nonumber
	\\
	\nonumber
	\Omega_H=&\frac{1}{2N} 
	\frac{u(2+s+2(1+s)u^2}{1+(1+s)u^2}
	\frac{\sqrt{s}}
	{\sqrt{(1+s)(1+u^2)(1-(1+s)u^4)}}\,.
\end{align} 
This solution has a variety of interesting properties, some of  them different from the case of 
 asymptotically  $\mathbb{M}^{1,4}$  Myers-Perry BHs (with the same symmetries). 
Here we mention only  the existence, for a given value of the $N$, of an upper bound of the spinning parameter $J$, with $J_{\rm max}=\frac{8 \pi N^3 }{G_5}$. 
For $J=J_{\rm max}$, the mass can take an arbitrary value $M>M_c=\frac{8\pi \sqrt{2} N^2}{G_5}$,
see Figure \ref{fig:7} (left panel). 
Note that the solution with $M=M_c$, $J=J_{\rm max}$ corresponds to an extremal  BH.
 Also, in the context of this work, it is interesting to consider the issue of solutions with constant $\Omega_H$. 
These sequences start at the horizonless GPS soliton limit;
however, their end point can be different. For $0<\Omega_H\leqslant 1/(\sqrt{2}N)$ they reach  an 
extremal BH solution ($T_H=0$) -
see the inset in Figure \ref{fig:7} (left panel); the extremal BH is marked there with a red dot. The situation is different for larger angular velocities,  
and a sequence of BHs at constant $ \Omega_H$ 
ends on the set of critical solutions with $J=J_{\rm max}$ and $M>M_c$.

\subsubsection{The extremal limit and an exact solution of the Klein-Gordon equation}
The extremal BH limit ($T_H=0$) is found for
\begin{eqnarray}
	s=\frac{1}{u^2}-1~,
\end{eqnarray} 
in which case the solution takes  a much simpler form. 
One finds
\begin{eqnarray}
	&&
	\nonumber
	{\cal F}_1= \frac{
		\left(1-\left(1-\frac{1}{u^2}\right)\frac{r_H}{r}\right)
		\left(1+\frac{1- u^2}{1+ u^2}\frac{r_H}{r}\right)
	}
	{\left(1-\frac{r_H}{r}\right)^2},~~
	{\cal F}_2= r^2 \left(1-\left(1-\frac{1}{u^2}\right)\frac{r_H}{r}\right)
	\left(1+\frac{1- u^2}{1+ u^2}\frac{r_H}{r}\right)),
	\\
	&&
	{\cal F}_3= \frac{r_H^2 
		\left(1+\frac{2r_H}{r}+\frac{5r_H^2}{r^2}
		+2\left(1-\frac{r_H}{r}\right)\left(1+\frac{5r_H}{r}\right)u^2
		+5 \left(1-\frac{r_H}{r}\right)^2 u^4
		\right)
	}
	{u^4\left(1+u^2\right)^2\left(1+\frac{1- u^2}{1+ u^2}\frac{r_H}{r}\right)\left(1-(1-\frac{1}{u^2}\right)\frac{r_H}{r})},
	\\
	&&
	\nonumber
	W=\frac{2u}{r}\sqrt{\frac{1+u^2}{1+2u^2+5 u^4}}
	\frac{2u^2\left(1+u^2\right)+\left(1-u^2\right)\left(1+2u^2\right)\frac{r_H}{r}}
	{ 1+\frac{2r_H}{r}+\frac{5r_H^2}{r^2}
		+2\left(1-\frac{r_H}{r}\right)\left(1+\frac{5r_H}{r}\right)u^2
		+5 \left(1-\frac{r_H}{r}\right)^2 u^4 }~,
\end{eqnarray} 
for the metric functions,
with the main quantities of interest are
\begin{eqnarray}
	&&
	M=\frac{4\pi N^2 }{G_5}\frac{\left(1+3 u^2\right)^3}{\left(1_2u^2+5u^4\right)^{3/2}}\,,~~~
	J=\frac{64 \pi N^3 u^3}{G_5}\left(\frac{1+u^2}{1+2u^2+5 u^4}\right)^{3/2}\,,
	\nonumber
	\\
	&&
	A_H=512 \pi^2 N^2 u^3 \left(\frac{1+u^2}{1+2u^2+5 u^4}\right)^{3/2},~~~
	\Omega_H=\frac{1}{4 N}\frac{1+3 u^2}{u \sqrt{1+u^2}},
	\\
	&&
	{\cal T}=\frac{N}{G_5} \frac{\left(1+3u^2\right)\left(1+3 u^4\right)}{\left(1+2u^2+5 u^4\right)^{3/2}}.
	\nonumber
\end{eqnarray} 
%

The Klein-Gordon equation (\ref{seq}) (with $s=1$)
 takes a relatively simple form
for the above (extremal) background 
\begin{eqnarray}
	\left(
	(r-r_H)^2 \phi'
	\right)'
	-\left( s_2 (r-r_H)^2+s_1 (r-r_H) +s_0 \right) \phi=0
\end{eqnarray} 
where we note
\begin{align}
\nonumber
	s_0 =\frac{3}{8}+\frac{2\mu^2 r_H^2}{u^2(1+u^2)}\,,\quad 
	s_1 =\left(\frac{\mu^2 r_H}{u^2(1+u^2)}-\frac{1}{16 r_H}\right)\left(1+3u^2\right)\,,~~
	s_2 =\mu^2-\frac{u^2\left(1+u^2\right)}{16r_H^2}\,.
\end{align} 
The general solution of the above equation reads
\begin{align}
	\nonumber
	\phi(r)=&\e^{-\sqrt{s_2} (r-r_H)}(r-r_H)^{\frac{1}{2} \left(\sqrt{1+4 s_0}-1\right)} \\
	&\times\bigg[
	c_1 L_{-\kappa}^{\sqrt{4 s_0+1}}\left(2 \sqrt{s_2} (r-r_H)\right)	\nonumber
	\\
	\label{sol-extremal}
	&\quad+c_2  U\left(\kappa,1+\sqrt{1+4 s_0},2 \sqrt{s_2} (r-r_H)\right)
	\bigg]
\end{align} 
where $U$ and $L$ are the  confluent hypergeometric function and the generalized Laguerre polynomial, respectively, and $\kappa = \frac{1}{2} \left(1+\sqrt{1+4 s_0}+\frac{s_1}{\sqrt{s_2}}\right)$ ($c_1, c_2$ arbitrary constants). This solution diverges either at spatial infinity or at the horizon.





\bibliographystyle{JHEP}
\bibliography{biblio}

\providecommand{\href}[2]{#2}\begingroup\raggedright\begin{thebibliography}{10}

\bibitem{kaluza}
T.~Kaluza, \emph{{Zum Unit\"atsproblem der Physik}}, \href{https://doi.org/10.1142/S0218271818700017}{\emph{Sitzungsber. Preuss. Akad. Wiss. Berlin (Math. Phys. )} {\bfseries 1921} (1921) 966} [\href{https://arxiv.org/abs/1803.08616}{{\ttfamily 1803.08616}}].

\bibitem{Klein:1926tv}
O.~Klein, \emph{{Quantum Theory and Five-Dimensional Theory of Relativity. (In German and English)}}, \href{https://doi.org/10.1007/BF01397481}{\emph{Z. Phys.} {\bfseries 37} (1926) 895}.

\bibitem{ACF}
T.~Appelquist, A.~Chodos and P.G.O.~Freund, eds., \emph{{Modern Kaluza-Klein theories}}, Frontiers in Physics, Addison-Wesley (1987).

\bibitem{Einstein:1943ixi}
A.~Einstein and W.~Pauli, \emph{{On the Non-Existence of Regular Stationary Solutions of Relativistic Field Equations}}, \href{https://doi.org/10.2307/1968759}{\emph{Annals Math.} {\bfseries 44} (1943) 131}.

\bibitem{Lichnerowicz}
A.~Lichnerowicz, \emph{{Sur le charact\`ere euclidien d'espaces-temps ext\'erieurs statiques partout r\'eguliers}}, {\emph{C. R. Hebd. Seanc. Acad. Sci.} {\bfseries 222} (1946) 432}.

\bibitem{Lichnerowicz1}
A.~Lichnerowicz, \emph{{Th\'eories relativistes de la gravitation et de l'Eelectromagn\'etisme}}, Paris, Masson (1955).

\bibitem{Gross:1983hb}
D.J.~Gross and M.J.~Perry, \emph{{Magnetic Monopoles in Kaluza-Klein Theories}}, \href{https://doi.org/10.1016/0550-3213(83)90462-5}{\emph{Nucl. Phys. B} {\bfseries 226} (1983) 29}.

\bibitem{Sorkin:1983ns}
R.D.~Sorkin, \emph{{Kaluza-Klein Monopole}}, \href{https://doi.org/10.1103/PhysRevLett.51.87}{\emph{Phys. Rev. Lett.} {\bfseries 51} (1983) 87}.

\bibitem{Hawking:1976jb}
S.W.~Hawking, \emph{{Gravitational Instantons}}, \href{https://doi.org/10.1016/0375-9601(77)90386-3}{\emph{Phys. Lett. A} {\bfseries 60} (1977) 81}.

\bibitem{Ishihara:2005dp}
H.~Ishihara and K.~Matsuno, \emph{{Kaluza-Klein black holes with squashed horizons}}, \href{https://doi.org/10.1143/PTP.116.417}{\emph{Prog. Theor. Phys.} {\bfseries 116} (2006) 417} [\href{https://arxiv.org/abs/hep-th/0510094}{{\ttfamily hep-th/0510094}}].

\bibitem{Yazadjiev:2006iv}
S.S.~Yazadjiev, \emph{{Dilaton black holes with squashed horizons and their thermodynamics}}, \href{https://doi.org/10.1103/PhysRevD.74.024022}{\emph{Phys. Rev. D} {\bfseries 74} (2006) 024022} [\href{https://arxiv.org/abs/hep-th/0605271}{{\ttfamily hep-th/0605271}}].

\bibitem{Brihaye:2006ws}
Y.~Brihaye and E.~Radu, \emph{{Kaluza-Klein black holes with squashed horizons and d=4 superposed monopoles}}, \href{https://doi.org/10.1016/j.physletb.2006.08.010}{\emph{Phys. Lett. B} {\bfseries 641} (2006) 212} [\href{https://arxiv.org/abs/hep-th/0606228}{{\ttfamily hep-th/0606228}}].

\bibitem{Nakagawa:2008rm}
T.~Nakagawa, H.~Ishihara, K.~Matsuno and S.~Tomizawa, \emph{{Charged Rotating Kaluza-Klein Black Holes in Five Dimensions}}, \href{https://doi.org/10.1103/PhysRevD.77.044040}{\emph{Phys. Rev. D} {\bfseries 77} (2008) 044040} [\href{https://arxiv.org/abs/0801.0164}{{\ttfamily 0801.0164}}].

\bibitem{Nedkova:2012yn}
P.G.~Nedkova and S.S.~Yazadjiev, \emph{{New Magnetized Squashed Black Holes -- Thermodynamics and Hawking Radiation}}, \href{https://doi.org/10.1140/epjc/s10052-013-2377-y}{\emph{Eur. Phys. J. C} {\bfseries 73} (2013) 2377} [\href{https://arxiv.org/abs/1211.5249}{{\ttfamily 1211.5249}}].

\bibitem{Chodos:1980df}
A.~Chodos and S.L.~Detweiler, \emph{{Spherically Symmetric Solutions in Five-dimensional General Relativity}}, \href{https://doi.org/10.1007/BF00756803}{\emph{Gen. Rel. Grav.} {\bfseries 14} (1982) 879}.

\bibitem{Dobiasch:1981vh}
P.~Dobiasch and D.~Maison, \emph{{Stationary, Spherically Symmetric Solutions of Jordan's Unified Theory of Gravity and Electromagnetism}}, \href{https://doi.org/10.1007/BF00756059}{\emph{Gen. Rel. Grav.} {\bfseries 14} (1982) 231}.

\bibitem{Pollard:1982gj}
D.~Pollard, \emph{{Antigravity and Classical Solutions of Five-dimensional {Kaluza-Klein} Theory}}, \href{https://doi.org/10.1088/0305-4470/16/3/015}{\emph{J. Phys. A} {\bfseries 16} (1983) 565}.

\bibitem{Gibbons:1985ac}
G.W.~Gibbons and D.L.~Wiltshire, \emph{{Black Holes in Kaluza-Klein Theory}}, \href{https://doi.org/10.1016/S0003-4916(86)80012-4}{\emph{Annals Phys.} {\bfseries 167} (1986) 201}.

\bibitem{Wang:2006nw}
T.~Wang, \emph{{A Rotating Kaluza-Klein black hole with squashed horizons}}, \href{https://doi.org/10.1016/j.nuclphysb.2006.09.001}{\emph{Nucl. Phys. B} {\bfseries 756} (2006) 86} [\href{https://arxiv.org/abs/hep-th/0605048}{{\ttfamily hep-th/0605048}}].

\bibitem{Bekenstein:1996pn}
J.D.~Bekenstein, \emph{{Black hole hair: 25 - years after}},  in \emph{{2nd International Sakharov Conference on Physics}}, pp.~216--219, 5, 1996 [\href{https://arxiv.org/abs/gr-qc/9605059}{{\ttfamily gr-qc/9605059}}].

\bibitem{Herdeiro:2014goa}
C.A.R.~Herdeiro and E.~Radu, \emph{{Kerr black holes with scalar hair}}, \href{https://doi.org/10.1103/PhysRevLett.112.221101}{\emph{Phys. Rev. Lett.} {\bfseries 112} (2014) 221101} [\href{https://arxiv.org/abs/1403.2757}{{\ttfamily 1403.2757}}].

\bibitem{Dias:2011at}
O.J.C.~Dias, G.T.~Horowitz and J.E.~Santos, \emph{{Black holes with only one Killing field}}, \href{https://doi.org/10.1007/JHEP07(2011)115}{\emph{JHEP} {\bfseries 07} (2011) 115} [\href{https://arxiv.org/abs/1105.4167}{{\ttfamily 1105.4167}}].

\bibitem{Herdeiro:2014ima}
C.A.R.~Herdeiro and E.~Radu, \emph{{A new spin on black hole hair}}, \href{https://doi.org/10.1142/S0218271814420140}{\emph{Int. J. Mod. Phys. D} {\bfseries 23} (2014) 1442014} [\href{https://arxiv.org/abs/1405.3696}{{\ttfamily 1405.3696}}].

\bibitem{Delgado:2016jxq}
J.F.M.~Delgado, C.A.R.~Herdeiro, E.~Radu and H.~Runarsson, \emph{{Kerr\textendash{}Newman black holes with scalar hair}}, \href{https://doi.org/10.1016/j.physletb.2016.08.032}{\emph{Phys. Lett. B} {\bfseries 761} (2016) 234} [\href{https://arxiv.org/abs/1608.00631}{{\ttfamily 1608.00631}}].

\bibitem{Brihaye:2014nba}
Y.~Brihaye, C.~Herdeiro and E.~Radu, \emph{{Myers\textendash{}Perry black holes with scalar hair and a mass gap}}, \href{https://doi.org/10.1016/j.physletb.2014.10.019}{\emph{Phys. Lett. B} {\bfseries 739} (2014) 1} [\href{https://arxiv.org/abs/1408.5581}{{\ttfamily 1408.5581}}].

\bibitem{Herdeiro:2015kha}
C.~Herdeiro, J.~Kunz, E.~Radu and B.~Subagyo, \emph{{Myers\textendash{}Perry black holes with scalar hair and a mass gap: Unequal spins}}, \href{https://doi.org/10.1016/j.physletb.2015.06.059}{\emph{Phys. Lett. B} {\bfseries 748} (2015) 30} [\href{https://arxiv.org/abs/1505.02407}{{\ttfamily 1505.02407}}].

\bibitem{Herdeiro:2017oyt}
C.~Herdeiro, J.~Kunz, E.~Radu and B.~Subagyo, \emph{{Probing the universality of synchronised hair around rotating black holes with Q-clouds}}, \href{https://doi.org/10.1016/j.physletb.2018.01.083}{\emph{Phys. Lett. B} {\bfseries 779} (2018) 151} [\href{https://arxiv.org/abs/1712.04286}{{\ttfamily 1712.04286}}].

\bibitem{Herdeiro:2016tmi}
C.~Herdeiro, E.~Radu and H.~R\'unarsson, \emph{{Kerr black holes with Proca hair}}, \href{https://doi.org/10.1088/0264-9381/33/15/154001}{\emph{Class. Quant. Grav.} {\bfseries 33} (2016) 154001} [\href{https://arxiv.org/abs/1603.02687}{{\ttfamily 1603.02687}}].

\bibitem{Santos:2020pmh}
N.M.~Santos, C.L.~Benone, L.C.B.~Crispino, C.A.R.~Herdeiro and E.~Radu, \emph{{Black holes with synchronised Proca hair: linear clouds and fundamental non-linear solutions}}, \href{https://doi.org/10.1007/JHEP07(2020)010}{\emph{JHEP} {\bfseries 07} (2020) 010} [\href{https://arxiv.org/abs/2004.09536}{{\ttfamily 2004.09536}}].

\bibitem{Hartmann:2010pm}
B.~Hartmann, B.~Kleihaus, J.~Kunz and M.~List, \emph{{Rotating Boson Stars in 5 Dimensions}}, \href{https://doi.org/10.1103/PhysRevD.82.084022}{\emph{Phys. Rev. D} {\bfseries 82} (2010) 084022} [\href{https://arxiv.org/abs/1008.3137}{{\ttfamily 1008.3137}}].

\bibitem{Liebling:2012fv}
S.L.~Liebling and C.~Palenzuela, \emph{{Dynamical boson stars}}, \href{https://doi.org/10.1007/s41114-023-00043-4}{\emph{Living Rev. Rel.} {\bfseries 26} (2023) 1} [\href{https://arxiv.org/abs/1202.5809}{{\ttfamily 1202.5809}}].

\bibitem{Herdeiro:2015gia}
C.~Herdeiro and E.~Radu, \emph{{Construction and physical properties of Kerr black holes with scalar hair}}, \href{https://doi.org/10.1088/0264-9381/32/14/144001}{\emph{Class. Quant. Grav.} {\bfseries 32} (2015) 144001} [\href{https://arxiv.org/abs/1501.04319}{{\ttfamily 1501.04319}}].

\bibitem{Hong:2019mcj}
J.-P.~Hong, M.~Suzuki and M.~Yamada, \emph{{Charged black holes in non-linear Q-clouds with O(3) symmetry}}, \href{https://doi.org/10.1016/j.physletb.2020.135324}{\emph{Phys. Lett. B} {\bfseries 803} (2020) 135324} [\href{https://arxiv.org/abs/1907.04982}{{\ttfamily 1907.04982}}].

\bibitem{Herdeiro:2020xmb}
C.A.R.~Herdeiro and E.~Radu, \emph{{Spherical electro-vacuum black holes with resonant, scalar $Q$-hair}}, \href{https://doi.org/10.1140/epjc/s10052-020-7976-9}{\emph{Eur. Phys. J. C} {\bfseries 80} (2020) 390} [\href{https://arxiv.org/abs/2004.00336}{{\ttfamily 2004.00336}}].

\bibitem{Hong:2020miv}
J.-P.~Hong, M.~Suzuki and M.~Yamada, \emph{{Spherically Symmetric Scalar Hair for Charged Black Holes}}, \href{https://doi.org/10.1103/PhysRevLett.125.111104}{\emph{Phys. Rev. Lett.} {\bfseries 125} (2020) 111104} [\href{https://arxiv.org/abs/2004.03148}{{\ttfamily 2004.03148}}].

\bibitem{Brihaye:2022afz}
Y.~Brihaye, C.~Herdeiro and E.~Radu, \emph{{$D = 5$ static, charged black holes, strings and rings with resonant, scalar Q-hair}}, \href{https://doi.org/10.1007/JHEP10(2022)153}{\emph{JHEP} {\bfseries 10} (2022) 153} [\href{https://arxiv.org/abs/2207.13114}{{\ttfamily 2207.13114}}].

\bibitem{Traschen:2001pb}
J.H.~Traschen and D.~Fox, \emph{{Tension perturbations of black brane space-times}}, \href{https://doi.org/10.1088/0264-9381/21/1/021}{\emph{Class. Quant. Grav.} {\bfseries 21} (2004) 289} [\href{https://arxiv.org/abs/gr-qc/0103106}{{\ttfamily gr-qc/0103106}}].

\bibitem{Townsend:2001rg}
P.K.~Townsend and M.~Zamaklar, \emph{{The First law of black brane mechanics}}, \href{https://doi.org/10.1088/0264-9381/18/23/320}{\emph{Class. Quant. Grav.} {\bfseries 18} (2001) 5269} [\href{https://arxiv.org/abs/hep-th/0107228}{{\ttfamily hep-th/0107228}}].

\bibitem{Brown:1992br}
J.D.~Brown and J.W.~York, Jr., \emph{{Quasilocal energy and conserved charges derived from the gravitational action}}, \href{https://doi.org/10.1103/PhysRevD.47.1407}{\emph{Phys. Rev. D} {\bfseries 47} (1993) 1407} [\href{https://arxiv.org/abs/gr-qc/9209012}{{\ttfamily gr-qc/9209012}}].

\bibitem{Kraus:1999di}
P.~Kraus, F.~Larsen and R.~Siebelink, \emph{{The gravitational action in asymptotically AdS and flat space-times}}, \href{https://doi.org/10.1016/S0550-3213(99)00549-0}{\emph{Nucl. Phys. B} {\bfseries 563} (1999) 259} [\href{https://arxiv.org/abs/hep-th/9906127}{{\ttfamily hep-th/9906127}}].

\bibitem{Lau:1999dp}
S.R.~Lau, \emph{{Light cone reference for total gravitational energy}}, \href{https://doi.org/10.1103/PhysRevD.60.104034}{\emph{Phys. Rev. D} {\bfseries 60} (1999) 104034} [\href{https://arxiv.org/abs/gr-qc/9903038}{{\ttfamily gr-qc/9903038}}].

\bibitem{Mann:1999pc}
R.B.~Mann, \emph{{Misner string entropy}}, \href{https://doi.org/10.1103/PhysRevD.60.104047}{\emph{Phys. Rev. D} {\bfseries 60} (1999) 104047} [\href{https://arxiv.org/abs/hep-th/9903229}{{\ttfamily hep-th/9903229}}].

\bibitem{Astefanesei:2005ad}
D.~Astefanesei and E.~Radu, \emph{{Quasilocal formalism and black ring thermodynamics}}, \href{https://doi.org/10.1103/PhysRevD.73.044014}{\emph{Phys. Rev. D} {\bfseries 73} (2006) 044014} [\href{https://arxiv.org/abs/hep-th/0509144}{{\ttfamily hep-th/0509144}}].

\bibitem{Henningson:1998gx}
M.~Henningson and K.~Skenderis, \emph{{The Holographic Weyl anomaly}}, \href{https://doi.org/10.1088/1126-6708/1998/07/023}{\emph{JHEP} {\bfseries 07} (1998) 023} [\href{https://arxiv.org/abs/hep-th/9806087}{{\ttfamily hep-th/9806087}}].

\bibitem{Balasubramanian:1999re}
V.~Balasubramanian and P.~Kraus, \emph{{A Stress tensor for Anti-de Sitter gravity}}, \href{https://doi.org/10.1007/s002200050764}{\emph{Commun. Math. Phys.} {\bfseries 208} (1999) 413} [\href{https://arxiv.org/abs/hep-th/9902121}{{\ttfamily hep-th/9902121}}].

\bibitem{Mann:2005cx}
R.B.~Mann and C.~Stelea, \emph{{On the gravitational energy of the Kaluza Klein monopole}}, \href{https://doi.org/10.1016/j.physletb.2006.02.025}{\emph{Phys. Lett. B} {\bfseries 634} (2006) 531} [\href{https://arxiv.org/abs/hep-th/0511180}{{\ttfamily hep-th/0511180}}].

\bibitem{Abbott:1981ff}
L.F.~Abbott and S.~Deser, \emph{{Stability of Gravity with a Cosmological Constant}}, \href{https://doi.org/10.1016/0550-3213(82)90049-9}{\emph{Nucl. Phys. B} {\bfseries 195} (1982) 76}.

\bibitem{Ortin:2004ms}
T.~Ortin, \emph{{Gravity and strings}}, Cambridge Monographs on Mathematical Physics, Cambridge Univ. Press (3, 2004), \href{https://doi.org/10.1017/CBO9780511616563}{10.1017/CBO9780511616563}.

\bibitem{Cai:2006td}
R.-G.~Cai, L.-M.~Cao and N.~Ohta, \emph{{Mass and thermodynamics of Kaluza-Klein black holes with squashed horizons}}, \href{https://doi.org/10.1016/j.physletb.2006.06.027}{\emph{Phys. Lett. B} {\bfseries 639} (2006) 354} [\href{https://arxiv.org/abs/hep-th/0603197}{{\ttfamily hep-th/0603197}}].

\bibitem{Peng:2017bjo}
J.-J.~Peng, \emph{{Revisiting the ADT mass of the five-dimensional rotating black holes with squashed horizons}}, \href{https://doi.org/10.1140/epjc/s10052-017-5290-y}{\emph{Eur. Phys. J. C} {\bfseries 77} (2017) 706}.

\bibitem{Hod:2012px}
S.~Hod, \emph{{Stationary Scalar Clouds Around Rotating Black Holes}}, \href{https://doi.org/10.1103/PhysRevD.86.129902}{\emph{Phys. Rev. D} {\bfseries 86} (2012) 104026} [\href{https://arxiv.org/abs/1211.3202}{{\ttfamily 1211.3202}}].

\bibitem{Hod:2013zza}
S.~Hod, \emph{{Stationary resonances of rapidly-rotating Kerr black holes}}, \href{https://doi.org/10.1140/epjc/s10052-013-2378-x}{\emph{Eur. Phys. J. C} {\bfseries 73} (2013) 2378} [\href{https://arxiv.org/abs/1311.5298}{{\ttfamily 1311.5298}}].

\bibitem{Benone:2014ssa}
C.L.~Benone, L.C.B.~Crispino, C.~Herdeiro and E.~Radu, \emph{{Kerr-Newman scalar clouds}}, \href{https://doi.org/10.1103/PhysRevD.90.104024}{\emph{Phys. Rev. D} {\bfseries 90} (2014) 104024} [\href{https://arxiv.org/abs/1409.1593}{{\ttfamily 1409.1593}}].

\bibitem{Cardoso:2005vk}
V.~Cardoso and S.~Yoshida, \emph{{Superradiant instabilities of rotating black branes and strings}}, \href{https://doi.org/10.1088/1126-6708/2005/07/009}{\emph{JHEP} {\bfseries 07} (2005) 009} [\href{https://arxiv.org/abs/hep-th/0502206}{{\ttfamily hep-th/0502206}}].

\bibitem{Kunduri:2006qa}
H.K.~Kunduri, J.~Lucietti and H.S.~Reall, \emph{{Gravitational perturbations of higher dimensional rotating black holes: Tensor perturbations}}, \href{https://doi.org/10.1103/PhysRevD.74.084021}{\emph{Phys. Rev. D} {\bfseries 74} (2006) 084021} [\href{https://arxiv.org/abs/hep-th/0606076}{{\ttfamily hep-th/0606076}}].

\bibitem{Myers:1986un}
R.C.~Myers and M.J.~Perry, \emph{{Black Holes in Higher Dimensional Space-Times}}, \href{https://doi.org/10.1016/0003-4916(86)90186-7}{\emph{Annals Phys.} {\bfseries 172} (1986) 304}.

\bibitem{schoen}
W.~Sch{\"o}nauer and R.~Wei\ss, \emph{{{\it The CADSOL Program Package,} Universit{\"a}t Karlsruhe, Interner Bericht Nr. 46/92 (1992)}}, {\emph{J. Comput. Appl. Math.} {\bfseries 27} (1989) 279}.

\bibitem{COLSYS}
U.~Ascher, J.~Christiansen and R.D.~Russel, \emph{{A collocation solver for mixed order systems of boundary value problems}}, {\emph{Math. of Comp.} {\bfseries 33} (1979) 659}.

\bibitem{Kaup:1968zz}
D.J.~Kaup, \emph{{Klein-Gordon Geon}}, \href{https://doi.org/10.1103/PhysRev.172.1331}{\emph{Phys. Rev.} {\bfseries 172} (1968) 1331}.

\bibitem{Ruffini:1969qy}
R.~Ruffini and S.~Bonazzola, \emph{{Systems of selfgravitating particles in general relativity and the concept of an equation of state}}, \href{https://doi.org/10.1103/PhysRev.187.1767}{\emph{Phys. Rev.} {\bfseries 187} (1969) 1767}.

\bibitem{Smarr:1972kt}
L.~Smarr, \emph{{Mass formula for Kerr black holes}}, \href{https://doi.org/10.1103/PhysRevLett.30.71}{\emph{Phys. Rev. Lett.} {\bfseries 30} (1973) 71}.

\bibitem{Bardeen:1973gs}
J.M.~Bardeen, B.~Carter and S.W.~Hawking, \emph{{The Four laws of black hole mechanics}}, \href{https://doi.org/10.1007/BF01645742}{\emph{Commun. Math. Phys.} {\bfseries 31} (1973) 161}.

\bibitem{Herdeiro:2014jaa}
C.~Herdeiro and E.~Radu, \emph{{Ergosurfaces for Kerr black holes with scalar hair}}, \href{https://doi.org/10.1103/PhysRevD.89.124018}{\emph{Phys. Rev. D} {\bfseries 89} (2014) 124018} [\href{https://arxiv.org/abs/1406.1225}{{\ttfamily 1406.1225}}].

\bibitem{Pena:1997cy}
I.~Pena and D.~Sudarsky, \emph{{Do collapsed boson stars result in new types of black holes?}}, \href{https://doi.org/10.1088/0264-9381/14/11/013}{\emph{Class. Quant. Grav.} {\bfseries 14} (1997) 3131}.

\bibitem{Herdeiro:2020kvf}
C.A.R.~Herdeiro, J.~Kunz, I.~Perapechka, E.~Radu and Y.~Shnir, \emph{{Multipolar boson stars: macroscopic Bose-Einstein condensates akin to hydrogen orbitals}}, \href{https://doi.org/10.1016/j.physletb.2020.136027}{\emph{Phys. Lett. B} {\bfseries 812} (2021) 136027} [\href{https://arxiv.org/abs/2008.10608}{{\ttfamily 2008.10608}}].

\bibitem{Harmark:2003dg}
T.~Harmark and N.A.~Obers, \emph{{New phase diagram for black holes and strings on cylinders}}, \href{https://doi.org/10.1088/0264-9381/21/6/026}{\emph{Class. Quant. Grav.} {\bfseries 21} (2004) 1709} [\href{https://arxiv.org/abs/hep-th/0309116}{{\ttfamily hep-th/0309116}}].

\bibitem{Harmark:2004ch}
T.~Harmark and N.A.~Obers, \emph{{General definition of gravitational tension}}, \href{https://doi.org/10.1088/1126-6708/2004/05/043}{\emph{JHEP} {\bfseries 05} (2004) 043} [\href{https://arxiv.org/abs/hep-th/0403103}{{\ttfamily hep-th/0403103}}].

\bibitem{Gervalle:2022npx}
R.~Gervalle and M.S.~Volkov, \emph{{Electroweak monopoles and their stability}}, \href{https://doi.org/10.1016/j.nuclphysb.2022.115937}{\emph{Nucl. Phys. B} {\bfseries 984} (2022) 115937} [\href{https://arxiv.org/abs/2203.16590}{{\ttfamily 2203.16590}}].

\bibitem{Gervalle:2022vxs}
R.~Gervalle and M.S.~Volkov, \emph{{Electroweak multi-monopoles}}, \href{https://doi.org/10.1016/j.nuclphysb.2023.116112}{\emph{Nucl. Phys. B} {\bfseries 987} (2023) 116112} [\href{https://arxiv.org/abs/2211.04875}{{\ttfamily 2211.04875}}].

\bibitem{Hod:2012wmy}
S.~Hod, \emph{{Stability of the extremal Reissner-Nordstroem black hole to charged scalar perturbations}}, \href{https://doi.org/10.1016/j.physletb.2012.06.043}{\emph{Phys. Lett. B} {\bfseries 713} (2012) 505} [\href{https://arxiv.org/abs/1304.6474}{{\ttfamily 1304.6474}}].

\bibitem{Hod:2013nn}
S.~Hod, \emph{{No-bomb theorem for charged Reissner-Nordstroem black holes}}, \href{https://doi.org/10.1016/j.physletb.2012.12.013}{\emph{Phys. Lett. B} {\bfseries 718} (2013) 1489}.

\bibitem{Jetzer:1989av}
P.~Jetzer and J.J.~van~der Bij, \emph{{CHARGED BOSON STARS}}, \href{https://doi.org/10.1016/0370-2693(89)90941-6}{\emph{Phys. Lett. B} {\bfseries 227} (1989) 341}.

\bibitem{Jetzer:1992tog}
P.~Jetzer, P.~Liljenberg and B.S.~Skagerstam, \emph{{Charged boson stars and vacuum instabilities}}, \href{https://doi.org/10.1016/0927-6505(93)90008-2}{\emph{Astropart. Phys.} {\bfseries 1} (1993) 429} [\href{https://arxiv.org/abs/astro-ph/9305014}{{\ttfamily astro-ph/9305014}}].

\bibitem{Pugliese:2013gsa}
D.~Pugliese, H.~Quevedo, J.A.~Rueda~H. and R.~Ruffini, \emph{{On charged boson stars}}, \href{https://doi.org/10.1103/PhysRevD.88.024053}{\emph{Phys. Rev. D} {\bfseries 88} (2013) 024053} [\href{https://arxiv.org/abs/1305.4241}{{\ttfamily 1305.4241}}].

\bibitem{Dowker:1994up}
F.~Dowker, J.P.~Gauntlett, S.B.~Giddings and G.T.~Horowitz, \emph{{On pair creation of extremal black holes and Kaluza-Klein monopoles}}, \href{https://doi.org/10.1103/PhysRevD.50.2662}{\emph{Phys. Rev. D} {\bfseries 50} (1994) 2662} [\href{https://arxiv.org/abs/hep-th/9312172}{{\ttfamily hep-th/9312172}}].

\bibitem{Dowker:1995gb}
F.~Dowker, J.P.~Gauntlett, G.W.~Gibbons and G.T.~Horowitz, \emph{{The Decay of magnetic fields in Kaluza-Klein theory}}, \href{https://doi.org/10.1103/PhysRevD.52.6929}{\emph{Phys. Rev. D} {\bfseries 52} (1995) 6929} [\href{https://arxiv.org/abs/hep-th/9507143}{{\ttfamily hep-th/9507143}}].

\bibitem{Rasheed:1995zv}
D.~Rasheed, \emph{{The Rotating dyonic black holes of Kaluza-Klein theory}}, \href{https://doi.org/10.1016/0550-3213(95)00396-A}{\emph{Nucl. Phys. B} {\bfseries 454} (1995) 379} [\href{https://arxiv.org/abs/hep-th/9505038}{{\ttfamily hep-th/9505038}}].

\bibitem{Matos:1996km}
T.~Matos and C.~Mora, \emph{{Stationary dilatons with arbitrary electromagnetic field}}, \href{https://doi.org/10.1088/0264-9381/14/8/027}{\emph{Class. Quant. Grav.} {\bfseries 14} (1997) 2331} [\href{https://arxiv.org/abs/hep-th/9610013}{{\ttfamily hep-th/9610013}}].

\bibitem{Larsen:1999pp}
F.~Larsen, \emph{{Rotating Kaluza-Klein black holes}}, \href{https://doi.org/10.1016/S0550-3213(00)00064-X}{\emph{Nucl. Phys. B} {\bfseries 575} (2000) 211} [\href{https://arxiv.org/abs/hep-th/9909102}{{\ttfamily hep-th/9909102}}].

\bibitem{Tomizawa:2008rh}
S.~Tomizawa and A.~Ishibashi, \emph{{Charged Black Holes in a Rotating Gross-Perry-Sorkin Monopole Background}}, \href{https://doi.org/10.1088/0264-9381/25/24/245007}{\emph{Class. Quant. Grav.} {\bfseries 25} (2008) 245007} [\href{https://arxiv.org/abs/0807.1564}{{\ttfamily 0807.1564}}].

\end{thebibliography}\endgroup
\end{document}